\documentclass[final]{mn2e}
\usepackage{umlaut}
\usepackage{color}
\usepackage{natbib}
\usepackage{graphicx}
\usepackage{graphics}
\usepackage{rotating}
\usepackage{amsmath}        
\usepackage{amssymb}

\date{This is a preprint of an Article accepted for publication in
  MNRAS. \copyright\ 2005 RAS}
\pagerange{\pageref{firstpage}--\pageref{lastpage}}
\pubyear{2005}

\def\LaTeX{L\kern-.36em\raise.3ex\hbox{a}\kern-.15em
    T\kern-.1667em\lower.7ex\hbox{E}\kern-.125emX}

\begin{document}

	\textheight 23cm

\newcommand{\Msun}{$M_{\odot}$}
\newcommand{\grs}{GRS\,1915+105}

\label{firstpage}

\title[Relativistic outbursts of GRS\,1915+105]{Multiple relativistic outbursts
of GRS\,1915+105: radio emission and internal shocks}

\author[J.C.A.~Miller-Jones et al.]
 {J.C.A.~Miller-Jones,$^1$\thanks{email: jmiller@science.uva.nl}
 D.G.~McCormick,$^2$ R.P.~Fender,$^3$ R.E.~Spencer,$^2$\and
 T.W.B.~Muxlow,$^2$ and G.G.~Pooley.$^4$\\
$^1$Astronomical Institute 'Anton Pannekoek', University of Amsterdam, Kruislaan  403,\\
1098 SJ, Amsterdam, The Netherlands.\\
$^2$The University of Manchester, Jodrell Bank Observatory, Cheshire,
SK11 9DL, UK\\
$^3$School of Physics and Astronomy, University of Southampton,
Highfield, Southampton, SO17 1BJ, UK\\
$^4$Mullard Radio Astronomy Observatory, Cavendish Laboratory,
 Madingley Road, Cambridge, CB3 0HE, UK}

\maketitle

\begin{abstract}

We present 5-GHz MERLIN radio images of the microquasar GRS\,1915+105
during two separate outbursts in 2001 March and 2001 July, following
the evolution of the jet components as they move outwards from the
core of the system.  Proper motions constrain the intrinsic jet speed
to be $>0.57c$, but the uncertainty in the source distance prevents an
accurate determination of the jet speed.  No deceleration is observed
in the jet components out to an angular separation of $\sim300$~mas.
Linear polarisation is observed in the approaching jet component, with
a gradual rotation in position angle and a decreasing fractional
polarisation with time.  Our data lend support to the internal shock
model whereby the jet velocity increases leading to internal shocks in
the pre-existing outflow before the jet switches off.  The compact
nuclear jet is seen to re-establish itself within two days, and is
visible as core emission at all epochs.  The energetics of the source
are calculated for the possible range of distances; a minimum power of
1--10 per cent $L_{\rm Edd}$ is required to launch the jet.

\end{abstract}


\section{Introduction.}

\grs\ was the first Galactic source observed to exhibit superluminal
motion \citep{Mir94}, and studies of its radio jets and jet-disc
interactions have been important in developing our understanding of
the link between accretion and jet outflows in X-ray binary systems.
The jets extract mass, energy, and angular momentum from the accreting
compact object, and in doing so can inject significant energy ($\geq
1$ per cent of the time-averaged luminosity of supernovae) into the ISM
(Fender, Maccarone \& van Kesteren 2005).

Jet outflows now seem to be ubiquitous in accreting systems such as
black hole X-ray binaries.  Such sources spend the majority of their
time in quiescence, with extremely low X-ray and radio luminosities
and a relatively hard X-ray spectrum.  In such states, they are
thought to be `jet-dominated', such that the power output in the jets
exceeds that radiated in X-rays (Fender, Gallo \& Jonker 2003).  As
the X-ray luminosity rises above $\sim7\times10^{-5}L_{\rm Edd}$, the
systems become X-ray dominated, with the majority of the accretion
energy being dissipated as X-rays in the inner parts of the accretion
flow.  The jets are then persistent, steady, and self-absorbed with a
flat radio spectrum (Stirling et al.\ 2001; Dhawan, Mirabel \&
Rodr\'\i guez 2000b).  At a high fraction ($\sim10^{-1}$) of the
Eddington luminosity, the central source makes a transition to a
softer X-ray state, passing through the so-called Very
High/Intermediate states.  As this happens, the jet velocity
increases, leading to internal shocks in the steady jet which appear
as highly relativistic knots moving away from the core of the system
\citep[e.g.][]{Fen99}.  At this stage, the steady-state jet outflow is
quenched until the system moves back into its low/hard X-ray state
once more.  This unified model is presented for generic black hole
X-ray binaries by Fender, Belloni \& Gallo (2004), and for the
specific case of \grs\ by \citet{Fen04a}.  We describe it in detail,
and in the light of the observations we present, in
\S~\ref{sec:model}.

The source \grs\ was discovered in 1992 via the {\sc watch} instrument
aboard the Russian {\sc granat} mission (Castro-Tirado, Brandt \& Lund
1992).  The system is believed to comprise a K-M \textsc{III} star
\citep{Gre01AA} and a $14\pm4 M_{\odot}$ black hole in a
$33.5\pm1.5$~d orbit (Greiner, Cuby \& McCaughrean 2001).  The nature
of the donor star implies that its mass should lie between 1 and
1.5~\Msun\ and rules out the stellar wind as the accretion mechanism;
the mass loss rate from such a star would be insufficient to feed the
high accretion luminosity of \grs. Thus it is thought that accretion
occurs via Roche lobe overflow.

In this paper, we present a study of the 2001 March and July outburst
of \grs, using data from the Rossi X-ray Timing Explorer ({\it RXTE})
and the Multi-Element Radio Linked Interferometer Network (MERLIN).
The X-ray observations are discussed in \S\,\ref{sec:xrays} and the
radio observations in \S\,\ref{sec:radio_obs}, which are used to place
constraints on the jet speed and inclination angle, the source
distance, and the possible deceleration and expansion of the jet knots
as they move outwards from the core.  The radio core emission is
analysed in \S\,\ref{sec:core}, and polarisation maps are presented in
\S\,\ref{sec:poln}.  We use the observations to constrain the
energetics of the source in \S\,\ref{sec:energetics}, and address the
discrepancy between radio observations of the jets in this source on
different angular scales in \S\,\ref{sec:previous}.

\section{X-ray States}
\label{sec:xrays}
The continuum X-ray spectrum of \grs, in common with most other black
hole sources, has been modelled as a superposition of a soft disc
blackbody and a hard power law \citep[e.g.][]{Kle02}. The blackbody
($k_{\rm B}T\sim$1--2\,keV) is associated with the accretion disc,
while the power law represents Comptonised emission from a corona,
which latter component may be ejected leading to a radio flare (or
oscillations in the case of repeated ejections).  The X-ray emission
is highly variable, which has been interpreted as the disappearance
and reappearance of the inner region of the accretion disc
\citep{Bel97a,Bel97b}, caused by the onset of thermal-viscous
instabilities.

\subsection{{\it RXTE} Data}

{\it RXTE}, launched in 1995, carries two pointed instruments, the
Proportional Counter Array (PCA) which covers the 2--60~keV energy
range, and the High Energy X-ray Timing Experiment (HEXTE) for the
15--250~keV energy range. Using archival PCA data \citet{Bel00} found
that the X-ray variability is the result of transitions between three
basic states; A, B, and C, categorised according to their spectral
hardness, temporal variability, and the prominence of the thermal disc
and power-law components.

\begin{figure}
\begin{center}
\includegraphics[width=0.45\textwidth,angle=0,clip=]{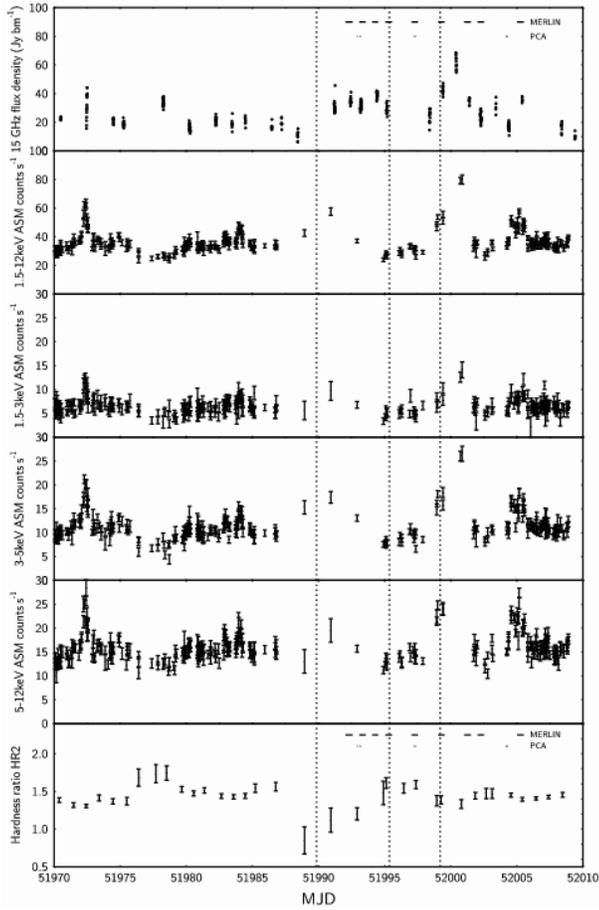}
\caption{15-GHz Ryle Telescope monitoring data, {\it RXTE} ASM count
  rates and hardness ratios prior to and during the 2001 March
  observations.  Vertical dotted lines show the derived
  zero-separation times for the three ejected components, and the top
  set of horizontal lines in the radio and HR$_2$ plots show the
  times of our MERLIN observations.  Below those are indicated the
  times of the {\it RXTE} PCA observations.  Note the spectral
  softening and peak in the count rate at the time of the first
  ejection event.}
\label{fig:march_xrays}
\end{center}
\end{figure}

\begin{figure}
\begin{center}
\includegraphics[width=0.45\textwidth,angle=0,clip=]{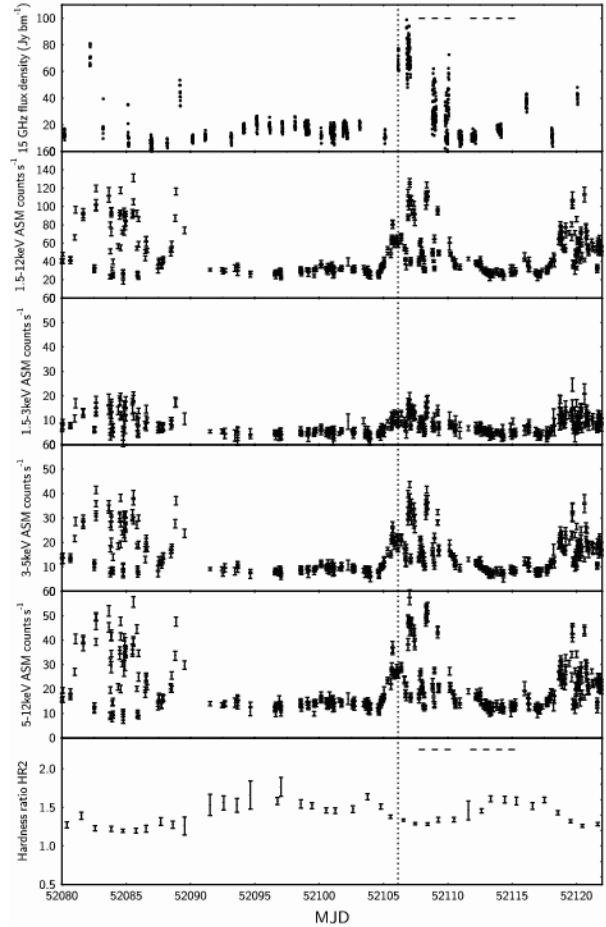}
\caption{15-GHz Ryle Telescope monitoring data, {\it RXTE} ASM count
  rates and hardness ratios prior to and during the 2001 July
  observations.  Vertical dotted lines show the derived
  zero-separation times for the ejected component, and the horizontal
  lines in the radio and HR$_{2}$ plots show the times of our MERLIN
  observations.}
\label{fig:july_xrays}
\end{center}
\end{figure}

The X-ray data presented here were collected by the All-Sky Monitor
(ASM) on board {\it RXTE}, which scans about 80 per cent of the sky
every orbit. The ASM returns the total source intensity in the
2--12\,keV band as well as intensities in three sub-bands: A
(1.3--3.0\,keV), B (3.0--5.0\,keV) and C (5.0--12.2\,keV). Two
hardness ratios can therefore be calculated; HR$_{1} = B/A$ and
HR$_{2}=C/B$. It should be noted that the bands and ratios for the PCA
differ due to that instrument's higher energy range (2--60\,keV) and
that PCA HR$_{1} \sim$ ASM HR$_{2}$.  The ASM count rates in the three
bands and the HR$_{2}$ hardness ratio are shown in
Figs.~\ref{fig:march_xrays} and \ref{fig:july_xrays} for the March and
July observations respectively.  In all cases, the radio outbursts
seem to be associated with a rise in the X-ray count rate and a degree
of spectral softening (a decrease in the hardness ratio HR$_{2}$).

PCA observations were made during the March observations at the epochs
indicated in Fig.~\ref{fig:march_xrays}, and in all cases, the
temporal variability and spectral hardness showed the source to be in
State C.  PCA observations were also made during the July outburst,
but since they have already been analysed by \citet{Vad03}, we will
not reanalyse them here.  They will be discussed in \S~\ref{sec:july}.

\section{Observations}
\label{sec:radio_obs}
\grs\ was observed with the MERLIN interferometer during two flaring
sequences, between 2001 March 24 and 2001 April 5, and also between
2001 July 17 and 2001 July 24.  The observing frequency was 4.994\,GHz
with a bandwidth of 15 MHz.  The observing times are listed in
Table\,\ref{tab:epochs}.  The March--April epochs of observation have
been labelled 1--9 sequentially, and the July observations labelled
10--16.  These target-of-opportunity observations were triggered by the
ongoing flux monitoring program at the Ryle Telescope \citep{Poo97}.
At each epoch, in addition to the target source \grs, observations
were made of a flux and polarisation angle calibrator, 3C\,286, a point
source calibrator, OQ\,208, B\,2134+004, or B\,0552+398, and a phase
reference source, B\,1919+086, 2.8$^{\circ}$ away from the target
source.  The point source calibrator used, and the time on source for
each epoch are also listed in Table\,\ref{tab:epochs}.  During the
observations, the five outstations (Cambridge, Defford, Knockin,
Darnhall and Tabley) were used, together with the Mark 2 antenna at
Jodrell Bank.

The MERLIN d-programs were used to perform initial data editing and
amplitude calibration, and the data were then imported into the
National Radio Astronomy Observatory's (NRAO) Astronomical Image
Processing System (\textsc{aips}) software package for further data
reduction.  The MERLIN pipeline was then used to image and
self-calibrate the phase reference source, and apply the derived
corrections to the target source, \grs.  The pipeline also calculated
the instrumental corrections (the \textit{D}-terms arising from signal
leakage from right circular polarisation feeds into left, and vice
versa) using B\,1919+086, and calibrated the polarisation position
angle using 3C\,286, assuming a position angle of 33$^{\circ}$ east of
north for its electric field vector.  Further self-calibration and
imaging were then carried out using standard procedures within
\textsc{aips}, using the phase-referenced images as initial models for
each epoch.  The solution interval was gradually reduced and the
number of \textsc{clean} components used in the model was increased
until there was no further improvement in the images.  The images in
Stokes \textit{I}, \textit{Q} and \textit{U}, were then combined to
produce images of total intensity, polarisation intensity, and
polarisation position angle.

Epochs 1, 6, 7 and 12 were all found to show significant core
variability on timescales of tens of minutes over the course of the
observing run.  This made imaging difficult, as the sidelobe levels
changed with time.  For epochs 1 and 7, the core flux densities were
approximately constant over a sufficiently long period of time that
imaging and self-calibration using only a restricted time range was
found to produce images with acceptably low residual r.m.s. noise
levels.  For epochs 6 and 12, the source flux densities were not
constant for long enough for this approach to work.  Instead, the data
were split into small segments (of length 20 and 30 minutes
respectively; a longer time period was required during the July
observations since the source was intrinsically fainter).  Each
segment was self-calibrated to remove any atmospheric jitter, imaged,
and the flux density of the core fitted and subtracted from the
\textit{uv}-data.  The segments were then recombined, and the
resulting core-subtracted data set was imaged to show any extension.
A point source was added back in at the fitted core position, with the
fitted mean flux density of the core, in order to make the final images
seen in Figs.\ \ref{fig:march_images}, \ref{fig:july_images} and
\ref{fig:poln}, and to enable easier comparisons to be made with the
other epochs.  The core parameters used were measured from the original
image of the entire dataset for that epoch.

\begin{table*}
\begin{center}
\caption{The labels, dates, Modified Julian Dates and on-source times for the
2001 MERLIN observations of \grs.}
\begin{tabular}{ccccc} \hline \hline
Epoch &  Date (2001)  &  MJD  & Time on source (min) & Point
source calibrator\\
\hline
1    &  March 24 & 51992.30225 $\pm$ 0.25878 & 478.5 & B\,0552+398\\
2    &  March 25 & 51993.32289 $\pm$ 0.23779 & 438.9 & B\,0552+398\\
3    &  March 26 & 51994.27843 $\pm$ 0.23574 & 433.7 & B\,0552+398\\
4    &  March 27 & 51995.29641 $\pm$ 0.25411 & 468.0 & B\,0552+398\\
5    &  March 29 & 51997.28781 $\pm$ 0.24570 & 431.3 & B\,2134+004\\ 
6    &  March 31 & 51999.28593 $\pm$ 0.24383 & 447.9 & B\,0552+398\\
7    &  April 2  & 52001.27083 $\pm$ 0.25383 & 468.0 & B\,0552+398\\
8    &  April 3  & 52002.27575 $\pm$ 0.25403 & 468.4 & B\,0552+398\\
9    &  April 6  & 52005.25733 $\pm$ 0.22043 & 408.4 & B\,0552+398\\
10   & July 17  & 52107.98903 $\pm$ 0.23861 & 372.1 & B\,0552+398\\
11   & July 18  & 52108.98906 $\pm$ 0.23865 & 369.7 & OQ\,208\\
12   & July 19  & 52109.98907 $\pm$ 0.23865 & 371.9 & B\,0552+398\\
13   & July 21  & 52111.98906 $\pm$ 0.23865 & 375.8 & OQ\,208\\
14   & July 22  & 52112.98282 $\pm$ 0.23239 & 357.7 & OQ\,208\\ 
15   & July 23  & 52113.98278 $\pm$ 0.23236 & 361.1 & OQ\,208\\
16   & July 24  & 52114.96038 $\pm$ 0.25060 & 388.3 & OQ\,208\\
\hline \hline
\end{tabular}
\label{tab:epochs}
\end{center}
\end{table*}

\subsection{March outburst}
\label{sec:march}

\begin{figure}
\begin{center}
\includegraphics[width=0.32\textwidth,angle=0,clip=]{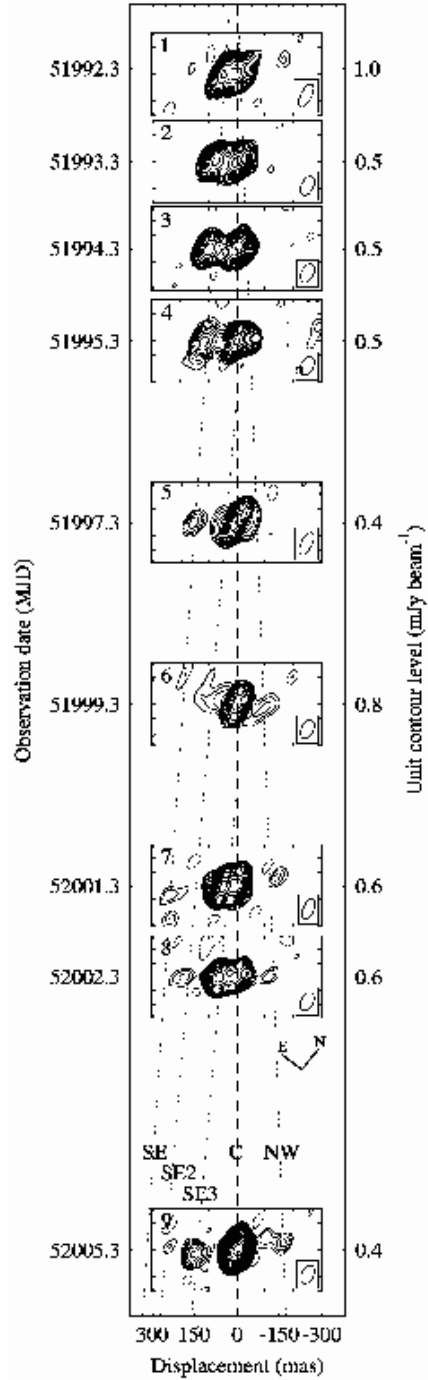}
\caption{Contour maps for the March observations. Solid and dashed
contours for each image are $(\sqrt{2}^{n})$ and $-(\sqrt{2}^{n})$
times the levels specified on the right-hand axis.  The images have
been rotated clockwise by $52.5^{\circ}$ to form the montage.  The
dotted lines correspond to the fitted ejection dates and proper
motions of the components.  The vertical dashed line indicates the
core position.  The beam sizes for each image are plotted in the lower
right-hand corner.  At a distance of $d_{\rm max}=10.9$~kpc, 1mas on
the image corresponds to a spatial scale of 10.9~au.  The component
labels are shown above the epoch 9 image, and the orientation on the
sky is also indicated.}
\label{fig:march_images}
\end{center}
\end{figure}

\begin{figure}
\begin{center}
\includegraphics[width=0.45\textwidth,angle=0,clip=]{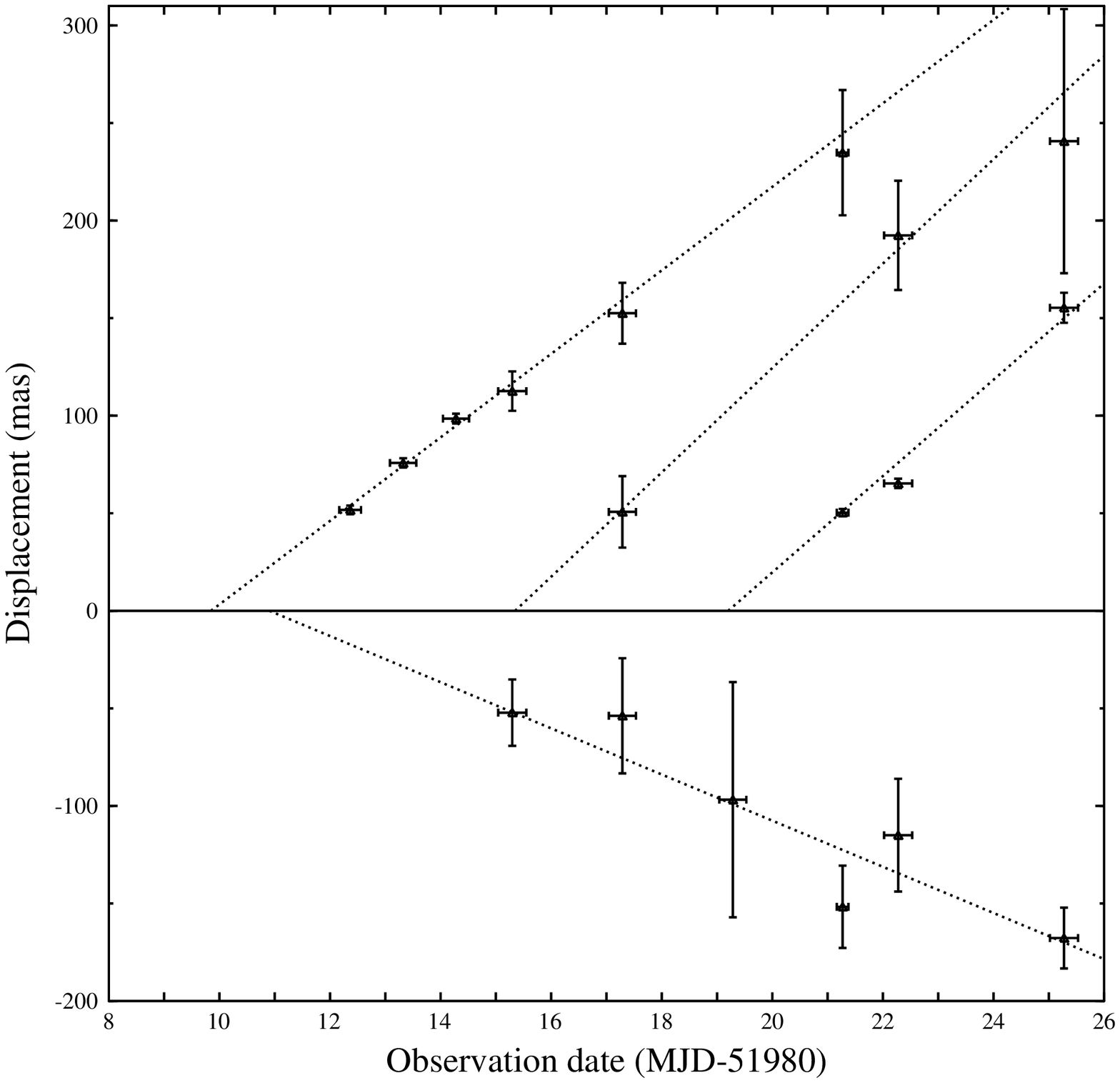}
\caption{Proper motions for the March observations.  The dotted lines
  correspond to the best fitting ejection dates and proper motions for
  the different components, accounting for uncertainties in both time
  of observation and measured angular separation.}
\label{fig:march_proper_motions}
\end{center}
\end{figure}

Fig.~\,\ref{fig:march_images} shows a composite of the MERLIN images
of the March observations.  The images have been rotated clockwise
through $52.5^{\circ}$, so that the southeastern components appear on
the left, and the northwestern components to the right.  The core
appears to be detected in all epochs (see \S~\ref{sec:core} for
further discussion), and we clearly observe components moving away to
the southeast and the northwest.  The southeastern components appear
to move faster (Fig.~\ref{fig:march_proper_motions}) and are brighter
at a given angular separation from the core
(Fig.~\ref{fig:flux_angsep}), and thus correspond to the approaching
jet, while the fainter, slower northwestern components are receding
from us.  This agrees with the findings of \citet{Mir94},
\citet{Rod99} and \citet{Fen99}.  Over the course of the 13 days of
observation, three distinct southeastern components were seen to be
ejected and move outwards, with the middle one being fainter than the
first and last.  These are labelled SE, SE2, SE3 respectively.  Owing
to Doppler deboosting of the receding jet flux density and its lower
apparent proper motion, only a single northwestern component was
observed, labelled NW.  The fitted flux densities and angular
separations from the core of the different components are given in
Table\,\ref{tab:march}.

Assuming that the ejecta move ballistically (i.e.\ with constant
velocity), straight-line fits to the angular separations can be
extrapolated to find the ejection dates of the different components.
These are given in Table~\ref{tab:proper_motions} and plotted in
Fig.~\ref{fig:march_proper_motions}, and reveal that components SE and
NW are consistent with having been ejected simultaneously.  The quoted
uncertainties take into account both the uncertainties in the fitted
positions of the core and the jet component, and also the
uncertainties in the time of observation, taken as half the length of
the observing run.  The Ryle telescope monitoring program
\citep{Poo97} indicates that the outburst in the 15-GHz flux density
of \grs\ which triggered our MERLIN observations peaked at MJD~51990.4
(see Fig.~\ref{fig:march_xrays}).  This, and also the observed sharp
rise in 15-GHz flux density, is in good agreement with our
zero-separation date of MJD~$51989.84\pm0.42$.
\begin{table*}
\begin{center}
\scriptsize
\caption{The fitted flux densities and angular separations from the core of
  components for the 2001 March observations of
  \grs.  The integrated flux densities and positions of the
  components were found by fitting Gaussians, and the angular
  separations were all measured from the fitted core positions in
  their respective images.}
\begin{tabular}{ccccccccccc} \hline \hline
Epoch & MJD & Core flux & SE sep. & SE flux & SE2 sep. &
SE2 flux & SE3 sep. & SE3 flux & NW sep. & NW flux\\
& -51990 & (mJy) & (mas) & (mJy) & (mas) & (mJy) & (mas) & (mJy) &
(mas) & (mJy)\\
\hline
1 & $2.36\pm0.20$ & $38.2\pm0.4$ & $51.7\pm2.2$ & $31.4\pm0.04$ & & &
& & &\\
2 & $3.32\pm0.24$ & $24.6\pm0.3$ & $75.8\pm2.4$ & $21.1\pm0.03$ & & &
& & &\\
3 & $4.28\pm0.24$ & $29.4\pm0.3$ & $98.5\pm2.5$ & $19.0\pm0.3$ & & & &
& &\\
4 & $5.30\pm0.25$ & $27.8\pm0.5$ & $112.6\pm10.1$ & $10.5\pm0.6$ & & &
& & $52.2\pm17.0$ & $2.8\pm0.4$\\
5 & $7.29\pm0.25$ & $19.4\pm0.2$ & $152.6\pm15.6$ & $1.6\pm0.2$ &
$50.7\pm18.3$ & $2.3\pm0.3$ & & & $53.8\pm29.5$ & $1.9\pm0.3$\\
6 & $9.29\pm0.24$ & $49.7\pm0.2$ & & & & & & & $96.8\pm60.3$ &
$4.9\pm1.2$\\
7 & $11.27\pm0.10$ & $26.9\pm0.4$ & $234.8\pm32.1$ & $2.1\pm0.5$ & & &
$50.3\pm1.9$ & $51.1\pm0.4$ & $151.7\pm21.1$ & $1.6\pm0.4$\\
8 & $12.28\pm0.25$ & $19.6\pm0.3$ & & & $192.4\pm28.0$ & $2.2\pm0.4$ &
$65.3\pm2.4$ & $26.4\pm0.3$ & $115.0\pm28.9$ & $1.5\pm0.3$\\
9 & $15.28\pm0.25$ & $94.3\pm0.3$ & & & $240.7\pm67.7$ & $0.6\pm0.3$ &
$155.3\pm7.7$ & $4.2\pm0.3$ & $167.7\pm15.6$ & $1.1\pm0.2$\\
\hline \hline
\end{tabular}
\label{tab:march}
\end{center}
\end{table*}

\subsection{July outburst}
\label{sec:july}

\citet{Vad03} carried out a detailed study of the X-ray and radio
state of \grs\ prior to and during the July outburst.  The outburst
followed a by now familiar pattern, with a $\sim60$~mJy 15-GHz radio
flare on MJD~51989.0 \citep{Vad03} preceding a radio-loud plateau state,
which lasted until the relativistic ejection.  They found that
according to the {\it RXTE} PCA observations, prior to July 16, the
source was in a hard state (the radio-loud C-state, C$_{\rm RL}$), in
the $\chi_{\rm RL}$ class, and then moved to a state of enhanced X-ray
emission, possibly due to a disturbed accretion disc.  They were
however unable to determine the exact time of the X-ray state change.
During the radio flare, they saw X-ray dips, corresponding to periods
of state A, during which the Comptonised spectral component of the
X-ray emission was absent.

\begin{figure}
\begin{center}
\includegraphics[width=0.34\textwidth,angle=0,clip=]{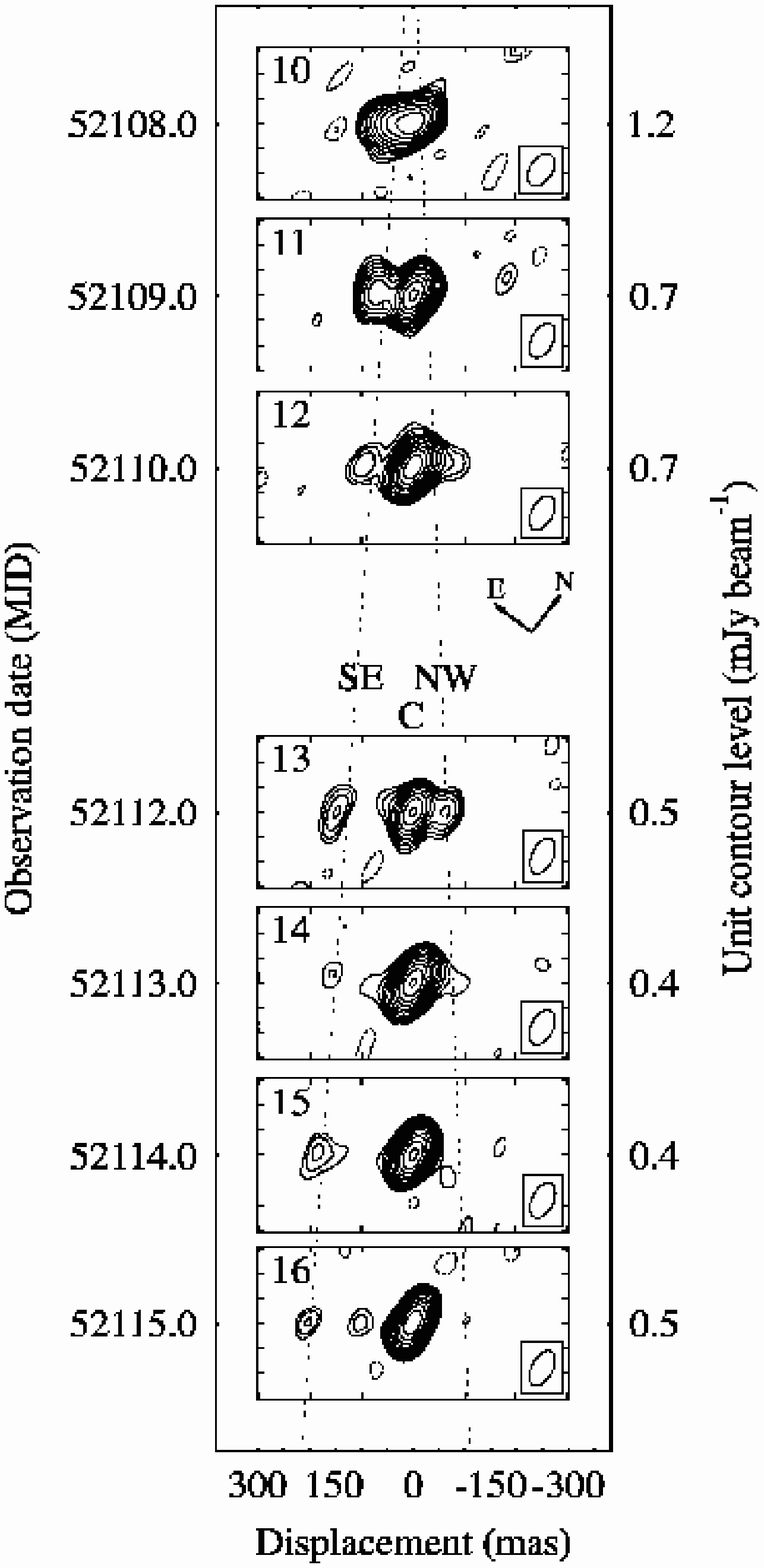}
\caption{Contour maps for the July observations. Solid and dashed
contours for each image are $(\sqrt{2}^{n})$ and $-(\sqrt{2}^{n})$
times the levels specified on the right-hand axis.  The images have
been rotated clockwise by $52.5^{\circ}$ to form the montage.  The
dotted lines correspond to the fitted ejection dates and proper
motions of the components.  The vertical dashed line indicates the
core position.  The beam sizes for each image are plotted in the lower
right-hand corner.  At a distance of $d_{\rm max}=10.9$~kpc, 1mas on
the image corresponds to a spatial scale of 10.9~au.}
\label{fig:july_images}
\end{center}
\end{figure}

During our July observations, from which the images are shown in
Fig.~\ref{fig:july_images}, the flux densities of the core and the jet
components were in general much lower than during the March
observations.  The flux densities of the core and SE component, and
the angular separation between the two, are given in
Table~\ref{tab:july}.  The only epochs in which an unambiguous
detection of the NW component was made were 12 and 13.  There is
marginal evidence in epoch 11 for an elongation of the core in the
opposite direction to that of the SE jet component, which could be
interpreted as the receding component.  There may be a north-western
extension to the core in epoch 14, though this is not particularly
convincing (having a flux density of only 1--2 times the r.m.s.\
noise).

\begin{figure}
\begin{center}
\includegraphics[width=0.45\textwidth,angle=0,clip=]{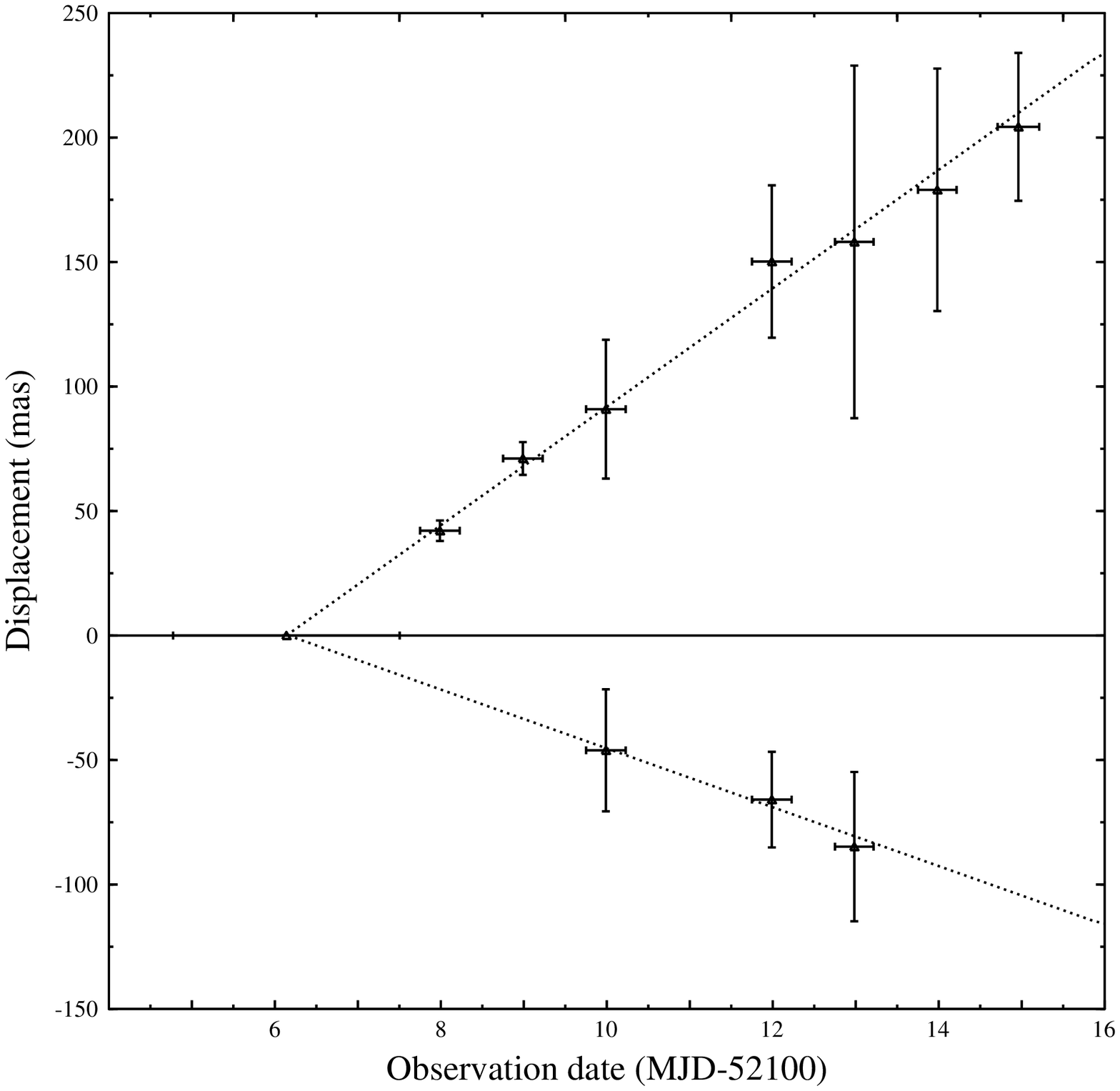}
\caption{Proper motions for the July observations.  The dotted lines
  correspond to the best fitting ejection dates and proper motions for
  the different components, accounting for the averaging time and the
  uncertainty in measured angular separation.}
\label{fig:july_proper_motions}
\end{center}
\end{figure}

Again assuming ballistic motion, a straight-line fit to the angular
separations of the SE component from the core gives a proper motion of
$\mu_{\rm app} = 23.8\pm2.7$~mas\,d$^{-1}$.  The fit is good, with
$\chi^{2}_{\rm red} = 0.4$, and implies a zero-separation date of
MJD~$52106.1 \pm 1.4$, which corresponds to July 16 03:21\,UT, in
good agreement with the start of the outburst determined by
\citet{Vad03}.  A similar fit to the receding component gives
$\mu_{\rm rec}=11.8\pm3.5$~mas\,d$^{-1}$, and a zero-separation date
in agreement with that for the approaching component.  These two fits
are shown in Fig.~\ref{fig:july_proper_motions}.

\begin{table*}
\begin{center}
\caption{The fitted flux densities and angular separations of components from
  the core for the 2001 July observations of \grs.}
\begin{tabular}{cccccccc} \hline \hline
Epoch & MJD-52100 & Angular separation & Error &  Core flux density & Error
& SE flux density & Error\\
& & (mas) & (mas) & (mJy) & (mJy) & (mJy) & (mJy)\\
\hline
10 & 7.99 & 42.1 & 4.1 & 28.0 & 0.6 & 33.6 & 0.8\\
11 & 8.99 & 71.1 & 6.6 & 27.1 & 0.4 & 8.9 & 0.4\\
12 & 9.99 & 90.9 & 27.9 & 27.7 & 0.5 & 2.6 & 0.5\\
13 & 11.99 & 150.2 & 30.6 & 9.6 & 0.4 & 2.0 & 0.4\\
14 & 12.98 & 158.1 & 70.8 & 15.1 & 0.3 & 0.6 & 0.3\\
15 & 13.98 & 179.0 & 48.7 & 15.1 & 0.3 & 1.2 & 0.4\\
16 & 14.96 & 204.3 & 29.7 & 16.3 & 0.3 & 0.8 & 0.3\\
\hline \hline
\end{tabular}
\label{tab:july}
\end{center}
\end{table*}

\begin{table*}
\begin{center}
\caption{Fitted proper motions and ejection dates of jet components,
  assuming ballistic motion}
\begin{tabular}{cccccc} \hline \hline
Outburst & Component & Epochs & Proper motion (mas\,d$^{-1}$) &
Ejection date (MJD) & $\chi^2_{\rm red}$\\
\hline
2001 March & SE & 1-5,7 & $21.4\pm2.0$ & $51989.84\pm0.42$ & 1.08\\
2001 March & NW & 4-9 & $11.8\pm2.0$ & $51990.9\pm2.2$ & 2.92\\
2001 March & SE2 & 5,8,9 & $26.8\pm5.9$ & $51995.4\pm2.6$ & 0.20\\
2001 March & SE3 & 7-9 & $27.4\pm2.5$ & $51999.6\pm1.6$ & 1.47\\
2001 March & SE3 & 6-9 & $24.7\pm1.0$ & $51999.2\pm0.6$ & 3.09\\
2001 July & SE & 10-16 & $23.8\pm2.8$ &$52106.1\pm1.4$ & 0.41\\
2001 July & NW & 0$^{a}$,12-14 & $11.8\pm3.5$ & $52106.2\pm3.8$ & 0.11\\
1997 Oct & SE & & $23.6\pm0.5$$^{b}$ & & $\leq 1$\\
1997 Oct & NW & & $10.0\pm0.5$$^{b}$ & & $\leq 1$\\
1994 Mar & SE & & $17.6\pm0.4$$^{c}$ & & $\leq 1$\\
1994 Mar & NW & & $9.0\pm0.1$$^{c}$ & & $\leq 1$\\
\hline
\multicolumn{6}{l}{$^{a}$ To better constrain the fit, the
  zero-separation date (`epoch 0') derived from the fit to the SE
  component was also used}\\
\multicolumn{6}{l}{$^{b}$ Data taken from \cite{Fen99}}\\
\multicolumn{6}{l}{$^{c}$ Data taken from \citet{Mir94}}\\
\hline\hline
\end{tabular}
\label{tab:proper_motions}
\end{center}
\end{table*}

\subsection{Proper motions}
\label{sec:proper_motions}
For a symmetric ejection event, if the proper motions of corresponding
approaching and receding components, $\mu_{\rm a}$ and $\mu_{\rm r}$
respectively, can be measured, it is possible to calculate the product
\begin{equation}
\beta\cos\theta = \frac{\mu_{\rm a}-\mu_{\rm r}}{\mu_{\rm a}-\mu_{\rm r}},
\end{equation}
where $\beta$ is the jet speed $v/c$ and $\theta$ is the inclination
angle of the jet axis to the line of sight.  Setting
$\theta=90^{\circ}$ allows us to place a lower limit, $\beta_{\rm
min}$ on the intrinsic jet velocity.  Assuming $\beta=1$ places an
upper limit, $\theta_{\rm max}$, on the inclination angle of the jet
axis to the line of sight, and allows an upper limit, $d_{\rm max}$,
to be placed on the source distance, given by
\begin{equation}
d_{\rm max} = \frac{c}{\sqrt{\mu_{\rm a}\mu_{\rm r}}}.
\end{equation}
For $\mu_{\rm a}$ and $\mu_{\rm r}$ in units of mas\,d$^{-1}$, this
may be more conveniently expressed as
\begin{equation}
d_{\rm max} = \frac{173}{\sqrt{\mu_{\rm a}\mu_{\rm r}}}\quad {\rm kpc}.
\end{equation}
Assuming that the SE and NW components for the March outburst were the
approaching and receding components from a single symmetric event,
then from the fitted proper motions given in
Table~\ref{tab:proper_motions}, we find $\beta\cos\theta =
0.29\pm0.09$.  This gives $\beta_{\rm min} = 0.29\pm0.09$,
$\theta_{\rm max} = 73.3\pm5.2^{\circ}$, and $d_{\rm max} =
10.9\pm1.0$~kpc.  If the distance is specified, then the exact values
of $\beta$ and $\theta$ can be calculated, as can the bulk Lorentz
factor $\Gamma = (1-\beta^2)^{-1/2}$, and the Doppler factors of the
approaching and receding jets, $\delta_{\rm app,rec} =
(\Gamma[1\mp\beta\cos\theta])^{-1}$.  These have been plotted in
Fig.~\ref{fig:jetparms}.

We can also use the proper motions from the July outburst given in
Table~\ref{tab:proper_motions} to find $\beta\cos\theta=0.34\pm0.14$,
$\theta_{\rm max} = 70.4^{\circ}\pm8.5^{\circ}$, and $d_{\rm max} =
10.3\pm1.6$~kpc.  These are consistent with the values found for the
March outburst, so for this reason, and since, owing to the more
robust detection of the receding component, the March values were
better-determined, the jet parameters for this outburst have not been
plotted.

Our derived proper motions are consistent (to within errors) with
those found by \citet{Fen99} with MERLIN and \citet{Dha00b} with the
VLBA, but greater than those found by \citet{Mir94} and \citet{Rod99}
with the VLA.  The implications of this are discussed further in
\S\,\ref{sec:previous}.

\subsection{Source distance}
\label{sec:distance}
\citet{Mir94} originally derived a distance of $12.5\pm1.5$\,kpc to
\grs\ from H\,\textsc{i} absorption measurements with the VLA, while
their measurements of jet knot proper motions constrained the maximum
possible source distance (corresponding to $\beta=1$) to be $d_{\rm
max}=13.7\pm0.2$\,kpc.  The proper motion observations of
\citet{Fen99} gave a somewhat smaller value, $d_{\rm
max}=11.2\pm0.8$\,kpc.  Dhawan, Goss \& Rodr\'\i guez (2000a) further
constrained the distance to \grs, again using H\,\textsc{i} absorption
measurements.  They found $d\geq 6.1$\,kpc and $d\leq12.2$\,kpc, and
adopted a distance of $12\pm1$\,kpc.  They also found extra absorption
along the line of sight to \grs\ compared with that to the
H\,\textsc{ii} region G\,45.46+0.06 (at a distance of 8.8\,kpc),
implying $d>9$\,kpc.  Recently however, \citet{Kai04} interpreted the
two \textit{IRAS} sources identified by \citet{Rod98} as the impact
sites of the jets of \grs\ on the interstellar medium.  This would
require all three sources to be at approximately the same distance
from us, $6\leq d \leq 7.4$\,kpc.  They also pointed out that the CO
observations of \citet{Cha04} allowed the constraint on the distance
being beyond the G\,45.46+0.06 to be relaxed.  \citet{Zdz05} dismissed
these assertions however, claiming that systematic errors on the
systemic velocity derived from infrared CO lines by \citet{Gre01Nat}
would be small.  The systemic velocity of \citeauthor{Gre01Nat}
together with the Galactic rotation curve gave a distance of
$12.1\pm0.8$~kpc.  Our 3$\sigma$ upper limits on $d_{\rm max}$ for the
March and July outbursts are 13.9 and 15.1~kpc respectively.

\begin{figure}
\begin{center}
\includegraphics[width=0.45\textwidth,angle=0,clip=]{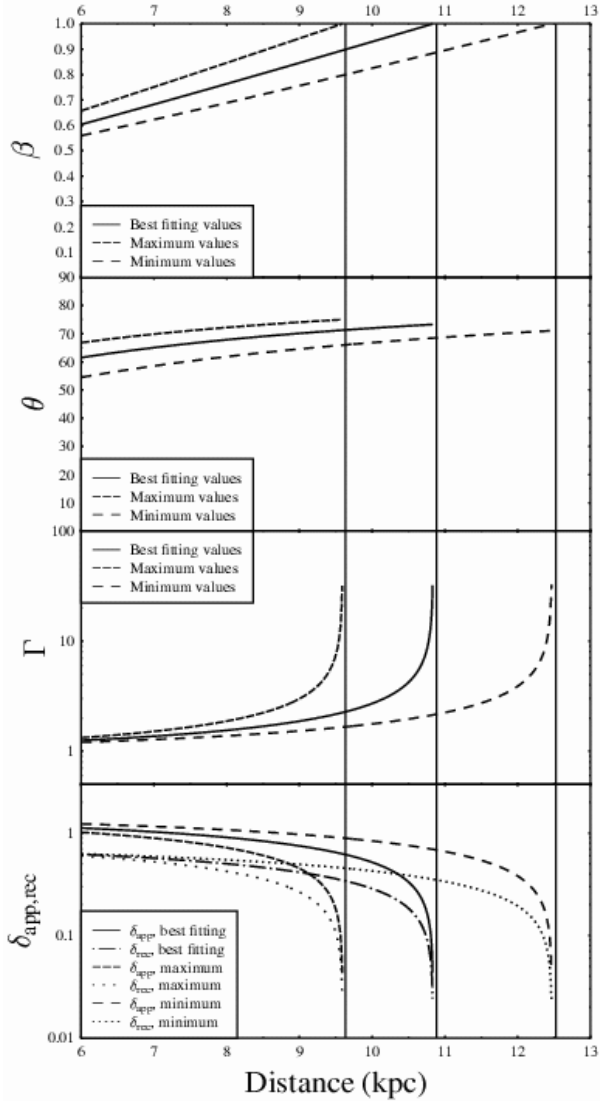}
\caption{Variation with assumed source distance of jet speeds,
  inclination angles, bulk Lorentz factors and Doppler factors for the
  approaching and receding jets.  Calculations have been done for the
  best fitting proper motions, the proper motions both low by
  $1\sigma$ (minimum values), and the proper motions both high by
  $1\sigma$ (maximum values).  The vertical lines indicate the values
  of $d_{\rm max}$, 10.9, 12.5, and 9.5~kpc respectively.}
\label{fig:jetparms}
\end{center}
\end{figure}

In summary, the distance would appear to lie between 6.1 and
12.2\,kpc, implying intrinsic jet speeds of $0.57<\beta<1$, from our
observations and Fig.~\ref{fig:jetparms}.  But it is not yet possible
to break the degeneracy between jet speed and inclination angle and
place any more rigorous constraints on the intrinsic jet speeds.
\citet{Dha00b} suggested that measurement of the annual trigonometric
parallax ($\sim 80$\,mas) might in future resolve this discrepancy.
Such a measurement would seem to be of prime importance in deciding
how relativistic the jets from this source actually are, and in
constraining the energy budget of the outbursts.

\citet{Fen03} remarked that all significantly relativistic jets lie
very close to $d_{\rm max}$ (which is in fact implicit in the
definition of $d_{\rm max}$ since it is the source distance
corresponding to $\beta=1$).  If the jets in \grs\ are as relativistic
as has previously been claimed \citep{Mir94,Fen99}, the source is
certainly at a distance of more than 9\,kpc.  Should it be as close as
6--7\,kpc, as postulated by \citet{Kai04}, the jets, while fast, would
not be the superluminal ejecta often assumed.

\subsubsection{Are the knots decelerating?}
\label{sec:deceleration}
As the jet knots move outwards, their bulk kinetic energy will be
thermalised as they interact with their environment.  Depending on the
rate at which this process occurs, it might be possible to measure the
deceleration of the jet knots, as has been done on arcsecond scales
for the case of XTE J\,1550--564 \citep{Cor02}.  The internal shock
model of Kaiser, Sunyaev \& Spruit (2000) however, predicts that for a
strong shock (which is believed to be the case), the velocity of the
shock would be constant.  In the case of \grs, \citet{Mir94} measured
significantly slower proper motions for both the approaching and
receding components than found here, at angular separations of up to
720~mas from the core.  However, their data were consistent with
ballistic motion, so deceleration would not seem to explain the
discrepancy.

Fitting our measured angular separations with a quadratic function,
rather than the straight-line fit required by ballistic motion, showed
no conclusive evidence for deceleration.  The best-fitting values for
the three components where there were a sufficient number of
measurements to properly constrain the fits (March SE, NW and July SE)
were decelerations of $0.7\pm0.5$, $0.4\pm0.8$, and
$0.9\pm1.3$~mas\,d$^{-2}$ respectively, all consistent with zero.  The
$\chi^2_{\rm red}$ values of these fits were similar to those for the
straight-line fits, and an \textit{F}-test \citep[e.g.][]{Pfe05}
showed that in no case was the probability greater than 95 per cent,
i.e.\ adding in the deceleration parameter was not necessary at any
significant level.  Furthermore, the quadratic fit to the angular
separation of the March SE component predicts an ejection date of
MJD~$51990.40\pm1.5$, the date of the 15-GHz flux density peak.
Assuming that the core flare corresponds to the ejection of that
component, the ejection date would then have been prior to the flux
density peak, strengthening the case against deceleration of that
component in our data.  These data do not therefore support the case
for deceleration.

\subsection{Expansion of jet components}
\label{sec:expansion}
\begin{figure}
\begin{center}
\includegraphics[width=0.45\textwidth,angle=0,clip=]{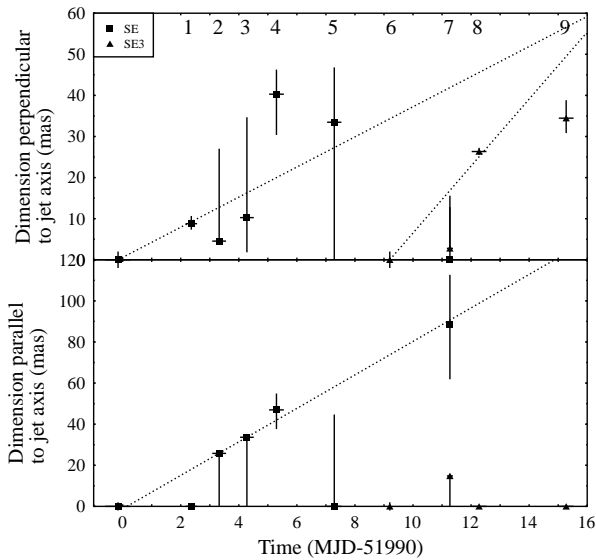}
\caption{Deconvolved component sizes for components SE and SE3 of the
  March observations, perpendicular (top) and parallel (bottom) to the
  jet axis.  Dotted lines show best-fitting linear expansions.  Labels
  denoting the epochs of observation are shown in the top plot.}
\label{fig:march_expansion}
\end{center}
\end{figure}

While no clear expansion of the jet component sizes is seen in the
images themselves, it should be remembered that the images of
Figs.~\ref{fig:march_images} and \ref{fig:july_images} are convolved
with the Gaussian restoring beam of $\sim 75\times40$~mas$^2$.
Deconvolving the beam from the fitted sizes to find the intrinsic
sizes of the Gaussian components fitted to the images shows marginal
evidence for the expansion of the jet knots.
Fig.~\ref{fig:march_expansion} shows fitted Gaussian knot sizes
parallel and perpendicular to the jet axis for components SE and SE3
of the March observations.
Component SE only shows convincing evidence for being resolved
perpendicular to the jet axis during epochs 1 and 5, whereas component
SE3 shows significant evidence for being resolved in both epochs 8 and
9.  The July observations were less conclusive, with the deconvolved
component sizes initially seeming to increase, but falling off again
in the later epochs as the flux density decreased and the knots became
harder to detect.

If this component expansion is real, it can be used to constrain
$\phi$, the opening angle of the jet.  We see the motion of the jet
knots projected on the plane of the sky.  This means that for a given
angular separation, $\delta\psi$, the true separation of the knot from
the core is $\delta\psi/\sin\theta$.  Thus, assuming spherical jet
knots,
\begin{equation}
\tan\phi = \frac{\dot{\zeta}\sin\theta}{\mu_{\rm app}},
\end{equation}
where $\zeta = r/d$ is the angular size of the source.  Without
knowing the exact source distance, we have only an upper limit to the
inclination angle of the jet axis to the line of sight.  Taking only
the most reliable point from the evolution of the SE component, that
of epoch 4, gives $\dot{\zeta} = 7.4\pm1.5$~mas\,d$^{-1}$, assuming
zero size at the ejection date derived in \S~\ref{sec:march}.
Alternatively, a linear fit to the data (shown as the dotted line in
Fig.~\ref{fig:march_expansion}) gives a value $\dot{\zeta} =
3.7\pm0.5$~mas\,d$^{-1}$.  These two values give upper limits on the
jet cone half-opening angle of $\phi_{\rm max} =
18.3^{\circ}\pm3.5^{\circ}$ and $9.4^{\circ}\pm1.3^{\circ}$ for the SE
component.  Since the receding component corresponding to knot SE3 is
not detected, $\theta_{\rm max}$ is not well-constrained in that case.

At a distance of $d_{\rm max}=10.9$~kpc, an angular expansion of
$7.4\pm1.5$~mas\,d$^{-1}$ corresponds to a lateral expansion speed
of $0.47\pm0.10c$.  We note that this is close to the theoretical
expansion speed of a relativistic plasma, $c/\sqrt{3} = 0.58c$.  The
data are unfortunately not good enough to make a more detailed
comparison possible in this case however.

\subsection{Flux densities}
\label{sec:flux_densities}
Accurately measuring the flux densities of the components is a
difficult procedure, since in order to adequately sample the
\textit{uv}-plane, and hence be able to detect the source structure, a
long integration is required.  But the jet components change in both
position and flux density with time.  During a 12-hour observation,
the proper motion of 21.4~mas\,d$^{-1}$ will cause a component to
move by 10.7~mas.

For the early epochs where the jet component was sufficiently bright
for its flux density to be determined accurately in a short time, the
core was subtracted from the data in the \textit{uv}-plane, and the
flux density at the position of the extension was measured as a
function of time using a direct Fourier transform of the complex
visibilities, binning the data into 6-minute time intervals.  At flux
densities below $\sim 5$~mJy\,beam$^{-1}$, the noise level in the data
made this impossible, and the flux density of the extension had to be
measured by fitting a Gaussian to the extension in the image plane.
The receding component was at no time sufficiently bright to use the
direct Fourier transform method.

Fitting a power-law decay with time to the measured flux densities for
components SE and SE3, and constraining the ejection time to be the
extrapolated time of zero-separation given in
Table~\ref{tab:proper_motions}, gave power-law indices of $1.8\pm0.03$
and $2.01\pm0.02$ respectively, with $\chi_{\rm red}^2$ values of 3.9
and 1.7 respectively.  The indices are Lorentz invariant, so are also
applicable to the flux density decay in the knot frame \citep{Rod99}.
Since the motion appears to be ballistic (\S~\ref{sec:deceleration}),
the flux density decays with angular separation with the same
power-law index.

\begin{figure}
\begin{center}
\includegraphics[width=0.45\textwidth,angle=0,clip=]{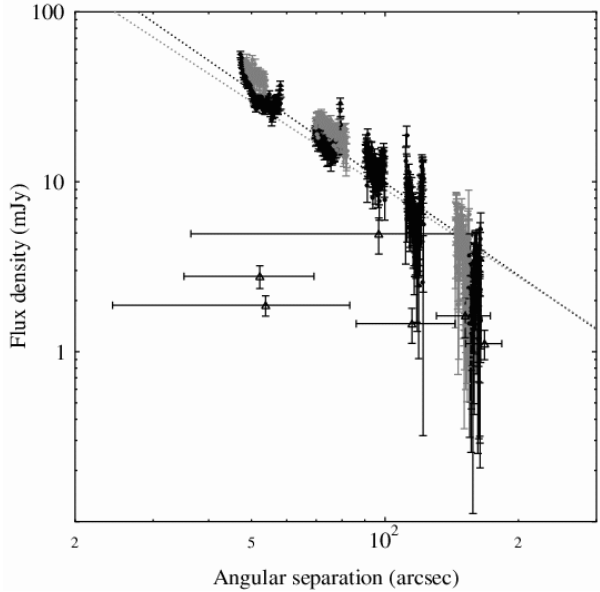}
\caption{Decrease in flux density with angular separation from the
  core, for the March observations.  Black points show the approaching
  component SE, grey points SE3, and open triangles the receding
  component NW.  Dotted black and grey lines show the best-fitting
  power law decays for components SE and SE3.  Flux densities for the
  approaching components were measured with time, converted to angular
  separations using the best-fitting proper motions for the two
  approaching components.}
\label{fig:flux_angsep}
\end{center}
\end{figure}

\subsubsection{Flux density ratios}
The ratio between the flux densities of approaching and receding
components, $S_{\rm app}$ and $S_{\rm rec}$ respectively, measured at
equal angular separation from the core, is given by
\begin{equation}
\frac{S_{\rm app}}{S_{\rm rec}} = \left(\frac{1+\beta\cos\theta}{1-\beta\cos\theta}\right)^{k+\alpha},
\end{equation}
where $\alpha$ is the spectral index of the emission, defined by
$S_{\nu} \propto \nu^{-\alpha}$, where $S_{\nu}$ is the flux density
at frequency $\nu$.  $k=2$ for a continuous jet, and $k=3$ for a jet
composed of discrete ejecta.  However, as pointed out by
\citet{Fen03}, a measurement of the flux density ratio gives no
information in addition to that already found from the proper motion
analysis of \S~\ref{sec:proper_motions}, other than the value of the
$k$-parameter.  Nevertheless, we can attempt to estimate this
parameter from the data.  The derived proper motions and
zero-separation dates of the approaching components were used to
convert the flux density decay with time found in
\S~\ref{sec:flux_densities} to flux density as a function of angular
separation from the core for the approaching component.  Making a
comparison to the flux density of the receding component should yield
a flux density ratio \textit{measured at equal angular separation}.
However, Fig.~\ref{fig:flux_angsep} shows that this is not a
straightforward procedure for these data.  At small angular
separations, where the flux density is high enough to be accurately
measured, but the angular separations are less accurate, the flux
density ratio is of order 10--15.  At greater angular separations, the
errors on the flux density determination are high, but the ratio is
closer to 4--5.  It appears that the flux density of the receding
component is still rising up to an angular separation of $\sim
100$~mas, although the uncertainties on the data are large, so the
flux density ratio is not constant as a function of angular
separation.  We now turn to a different approach to measuring $k$.

Miller-Jones, Blundell \& Duffy (2004) derived an expression relating
the value of $\beta\cos\theta$ to the flux density ratio in a single
image, assuming adiabatic expansion of the jet components, with the
knot radius scaling linearly with time.  In such a case, the flux
density ratio \textit{in a single image} is given by
\begin{equation}
\frac{S_{\rm app}}{S_{\rm rec}} =
\left(\frac{1+\beta\cos\theta}{1-\beta\cos\theta}\right)^{k-p},
\label{eq:flux_density_ratio}
\end{equation}
where $p$ is the index of the electron spectrum, such that $N(E) =
\kappa E^{-p}$.  We note however that if the rise phase was optically
thin, and the image used was made at a time when the flux density of
the receding jet was still increasing while that of the approaching
jet was decreasing owing to adiabatic expansion, then this formalism
would be invalid, since both knots must be in the adiabatic expansion
regime of decreasing flux density.  As already remarked,
Fig.~\ref{fig:flux_angsep} appears to show that the flux density of
the receding knot is still rising up to an angular separation of
$\sim100$~mas (with a large error bar), and only after that does it
start decaying, so care must be taken when applying the above
formalism.

The only epochs for which corresponding approaching and receding jets
were both clearly visible and distinguishable from the core were 5, 7,
and 13.  These gave flux density ratios of $0.89\pm0.17$,
$1.27\pm0.47$, and $0.76\pm0.19$ respectively.  We note that in two of
the three cases, the receding jet is brighter than the approaching
jet, and the third case is in fact consistent with this scenario.
This would constrain $k<p$.  Since we do not have multiwavelength
observations, it is not possible to deduce the spectral index of the
jet knots, and hence the value of $p = 2\alpha+1$.  However,
\citet{Fen02} observed this same outburst at both 4.8 and 8.64~GHz,
and found a spectral index for the integrated (core plus jet knots)
flux density of the source which stabilised at $\sim0.77\pm0.05$ by
MJD~51991.9.  \citet{Mir94} measured a spectral index of
$0.84\pm0.03$ for both the approaching and the receding jet components
(once they had separated from the core) in the outburst of 1994 March.
Assuming that the spectral index was the same in the outbursts we
observed, this yields $k=2.5\pm0.3$ for epoch 5, $k=3.1\pm0.6$ for
epoch 7, and $k=2.3\pm0.4$ for epoch 13.  These are significantly
greater than the values of 1.3--1.9 found by \citet{Fen99}, but
consistent with the value of 2.3 found by \citet{Mir94}, and would
seem to imply something intermediate between discrete ejections and a
steady jet.  The high value for epoch 7 would seem to argue for a
discrete ejection.  However, it would seem strange for the character
of the outburst (i.e.\ the value of $k$) to change as the knots moved
outwards.  A possible reason for the discrepancy could be that the
formula was not applicable during epochs 5 and 13.  The angular
separation of the receding component during epoch 5 is
$53.8\pm29.5$~mas which, from Fig.~\ref{fig:flux_angsep}, would imply
that $k$ cannot be calculated accurately for this epoch using
Equation~\ref{eq:flux_density_ratio}, thus lending more credibility to the
value derived for epoch 7.  Since we have no similar plot for the July
outburst, it is not possible to evaluate whether or not such an effect
could also explain the low value of $k$ for epoch 13.

\section{The core}
\label{sec:core}
\begin{figure}
\begin{center}
\includegraphics[width=0.45\textwidth,angle=0,clip=]{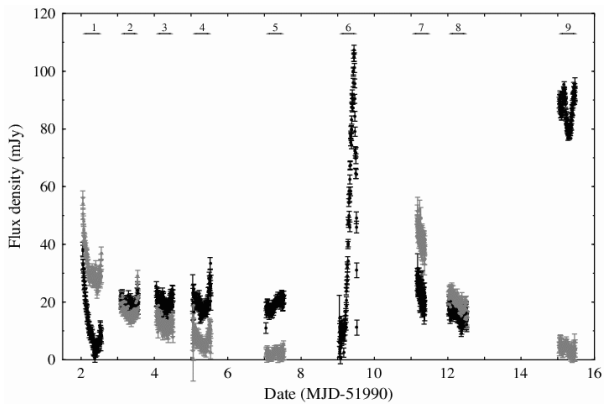}
\caption{Flux density of the core during the 2001 March observations.
The time range of each observation epoch is indicated at the top of
the plot.  Epochs 1, 6 and 7 all show a fractional variation of at
least 50 per cent.  The black points are for the core, and the grey points
are for the approaching component, SE for epochs 1--5, and SE3 for
epochs 7--9.}
\label{fig:core_flux}
\end{center}
\end{figure}

Detectable emission seems to be associated with what we interpret as
the core of the system at all epochs, although in some of the
observations it appears to be blended with an emerging approaching or
receding jet component.  This is in contrast to the observations of
\citet{Fen99} in which the flux density was always dominated by the
jet components.  Fig.~\ref{fig:core_flux} shows the flux density of
the core and jet components during the March observations.  As will be
explained further in \S\,\ref{sec:poln}, the core was not in general
found to show significant linear polarisation, except possibly for
epochs 2 and 9 of the March outburst sequence.

Epochs 1 and 6 show that the core flux density faded to very low
levels immediately after a flare.  This can be identified as the jet
suppression in the soft disc-dominated state in the model of
\citet{Fen04b} (see \S~\ref{sec:model} for details).  However, by the
start of epoch 2, the core flux density had recovered and stabilised
at a level of approximately 20~mJy.  We attribute this low-level radio
emission to the steady, compact nuclear jet.  This stable level of
$\sim20$~mJy at 5~GHz was also seen following the July outburst.  This
implies that the source has moved back to a harder X-ray state to the
right of the jet-line (see \S~\ref{sec:model} for further details)
{\it without} launching a second major ejection.  Only crossing the
line from right to left in fig.~7 of \citet{Fen04b} (hard to soft
X-ray state) gives rise to the internal shocks and the corresponding
relativistic ejecta.

\begin{figure}
\begin{center}
\includegraphics[width=0.45\textwidth,angle=0,clip=]{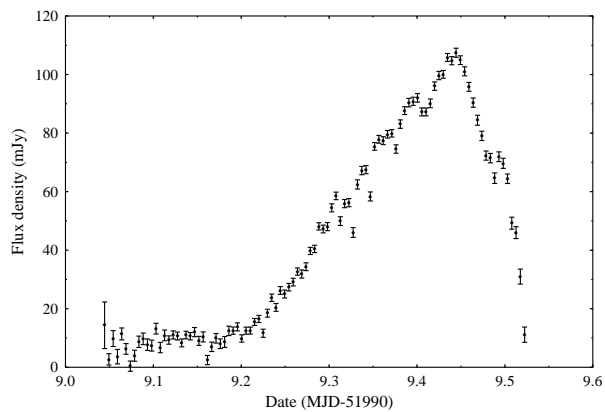}
\caption{Flux density of the core during epoch 6 of the 2001 March
observations.  This does not follow the fast rise and exponential
decay (FRED) behaviour observed for the integrated flux density of
such systems.  Note the rise from and fall back to very low flux
densities ($\leq10$~mJy) either side of the core flare.}
\label{fig:core_flux6}
\end{center}
\end{figure}

The flare in core flux density seen during epoch 6 of the March
observations (Fig.~\ref{fig:core_flux6}) corresponds to the ejection
of component SE3.  The core flux density began to rise from a flat
base level of $9.8\pm1.2$~mJy at MJD~$51999.20\pm0.02$, peaking at
$107.4\pm1.4$~mJy 0.24~d later.  Assuming the start of the rise phase
to be the ejection date of the component, we get an extra constraint
on the proper motion of component SE3, used in the fit for epochs 6--9
given in \S~\ref{sec:july} and Table~\ref{tab:proper_motions}.

A detailed inspection of Fig.~\ref{fig:core_flux6} seems to show
evidence for slightly more variable radio emission immediately prior
to the rise in flux density corresponding to the flare, as originally
noted by \citet{Fen04b}.  But we note that this, together with the
rapid falloff in flux density at the very end of the observing run,
occurred when the source elevation was low ($\lesssim 15^{\circ}$), at
which point the weather begins to affect the gain solutions (otherwise
good to $\sim 5$ per cent).  A check on the flux density of the phase
calibrator confirmed that only for the time range covering the last
two points in Fig.~\ref{fig:core_flux6} were the data thus affected.
The rest of the data show that the core jet flux density decreased
with an e-folding time of $\sim0.07$\,d.  The data from epoch 1 show a
very low core flux density (observed when the source was still at a
fairly high elevation ($\sim 35^{\circ}$)).  This is $2.5\pm0.4$~d
after the derived ejection date of the SE component, by which time the
jet knot would have moved out to an angular separation of
$\sim55$~mas, i.e.\ out of the beam.  We cannot accurately probe the
core flux density on its own until this time, as we will be unable to
decouple the core emission from the jet knot emission.  But the low
flux density seen during epoch 1 suggests that the steady core jet is
not re-established until the increase in core flux density 2.6~d after
the ejection event.  This is considerably longer than the timescale of
18~h found by \citet{Dha00b} for the nuclear jet to reform following
the start of a major outburst, derived from the date of the first
observations \citep{Fen99} of radio oscillations following the
outburst.

No short-period (20--40\,min) oscillations of the core flux density
such as those described by \citet{Poo97} were observed.  A search was
made by measuring the flux density in the \textit{uv}-plane at the
core position in 1-minute time bins, having subtracted off any
extension from the data.  The resulting time series was run through
one-dimensional CLEAN algorithm (Roberts, Leh\'ar \& Dreher 1987),
which showed no significant peaks in the power spectrum of the data on
frequencies shorter than 2\,d$^{-1}$.

The flux density of the core was much higher during epoch 9, at round
80-100~mJy, and seemed to oscillate with a period of approximately 8
hours.  We cannot however determine whether these are true
oscillations, since we only sample $\sim1.5$ of these `periods' in a
12-hour observing run.

\subsection{Positional accuracy}
The beam sizes for the various images are shown in
Figs.~\ref{fig:march_images} and \ref{fig:july_images}.  But the
use of phase referencing allows the determination of the target
position to much greater accuracy, set by the uncertainty in the
position of the phase calibrator, the uncertainty in the telescope
positions, and atmospheric phase winding.  B\,1919+086 has a position
accurate to $\approx 14$\,mas \citep{Bro98}, and being 2.84$^{\circ}$
from \grs, introduces a positional uncertainty of about 10\,mas if
the atmosphere causes phase wraps on timescales of about 2 hours (the
worst-case scenario in the data).

The positions of the core component were measured prior to
self-calibration for each epoch, and a weighted mean position was
taken for the July observations.  This gave a B\,1950 position of
$19^{\rm h} 12^{\rm m} 49^{\rm s}.96941\pm0^{\rm s}.00009$,
$10^{\circ} 51^{\prime} 26^{\prime\prime}.6427\pm0.0051$, with a
spread of 0.25\,ms in RA and 15\,mas in Dec.  Owing to the multiple
ejections during the March observations, the core position was harder
to measure accurately, owing to the blending of the core with
newly-emerging components.  This lead to a much greater spread in the
measured core co-ordinate, of 2.67\,ms in RA (1.12\,ms neglecting
epoch 7) and 17.1\,mas in Dec.  The weighted mean position was
$19^{\rm h} 12^{\rm m} 49^{\rm s}.96945\pm0^{\rm s}.00053$,
$10^{\circ} 51^{\prime} 26^{\prime\prime}.6588\pm0.0060$ (B\,1950).  We
note that these positions all rely on having correctly identified the
relatively stationary central component in the images as the core.
Furthermore, the systematic errors in the astrometry mentioned in the
previous paragraph mean that the uncertainty on these derived absolute
positions is at least of order 15\,mas, although we did not have an
astrometric check source we could use to quantify accurately this
uncertainty.  Since the hour angle coverage was approximately the same
within the each set of observations (March and July), and since the
same phase calibrator source was used in all observations, the errors
in the relative positions from epoch to epoch are much smaller, and
are dominated mainly by the signal-to-noise ratio and the atmospheric
phase winding.  The major source of error in the angular separations
within a single image is the signal-to-noise ratio of the knots.

The most accurate determination of the position of the core of \grs\
is currently \citep{Dha00b} 19$^{\rm h}$15$^{\rm m}$11.$^{\rm
s}$54938$\pm0.^{\rm s}00007$,
10$^{\circ}$56$^{\prime}$44.$^{\prime\prime}7585\pm0.^{\prime\prime}001$
(J\,2000) on 1998 May 2.  The proper motion on the sky was determined
as $5.8\pm1.5$\,mas\,yr$^{-1}$, ascribed to secular parallax.  In the
three months separating our sets of observations, this corresponds to
a shift of $1.8\pm0.5$\,mas between our March and July observations.
Our uncertainties are not sufficiently small to be able to verify this
figure.  However, comparison between our position and that of
\citet{Dha00b} is difficult, since the two sets of observations used
different calibrators for the phase referencing, and since MERLIN uses
B\,1950 co-ordinates whereas the VLBA uses J\,2000 co-ordinates.

\section{Polarisation}
\label{sec:poln}
\subsection{Linear polarisation}
\begin{figure}
\begin{center}
\includegraphics[height=0.8\textheight,angle=0,clip=]{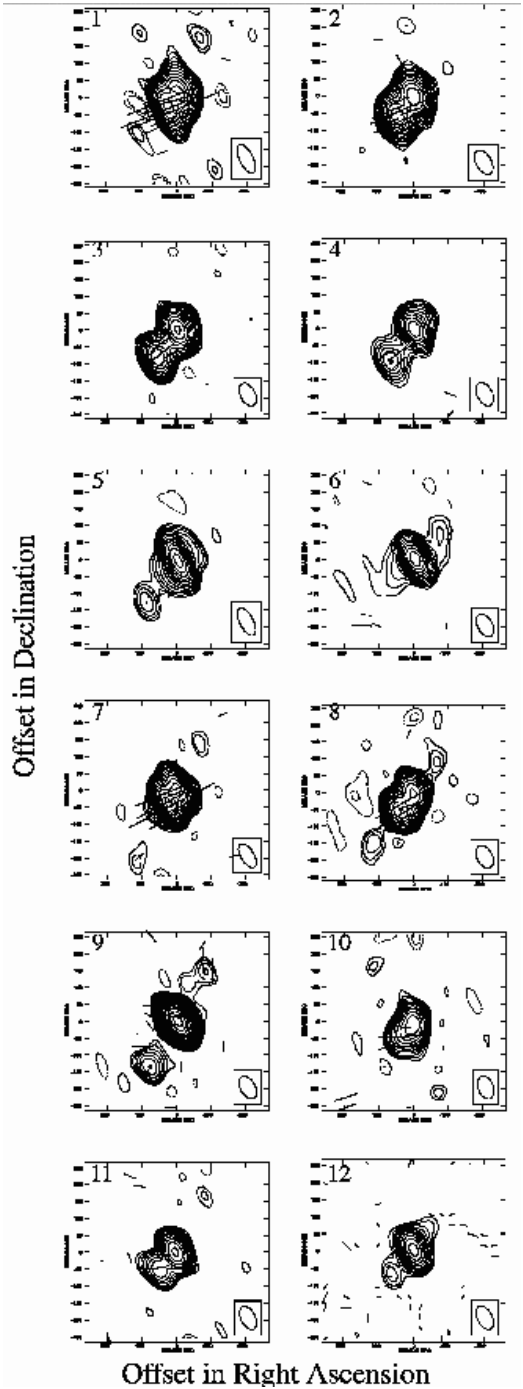}
\caption{Contour maps with superposed linear polarisation $E$
vectors. Solid and dashed contours are $(\sqrt{2})^{n}$ and
$-(\sqrt{2})^{n}$ times the 3-$\sigma$ noise level in the image.  The
lengths of the polarisation vectors are scaled to represent the level
of linearly polarised flux density $P$ in the image.  A vector of
length 100~mas corresponds to 3.33~mJy\,beam$^{-1}$.  To reduce
confusion, only polarisation vectors greater than the 3-$\sigma$ level
in the total polarisation image have been plotted, except in the case
of epoch 12, where the vectors have been plotted down to the 1-$\sigma$
level.  The beam sizes for each image are plotted in the lower
right-hand corner.  At a distance of 10.9~kpc, 1mas on the image
corresponds to a spatial scale of 10.9~au.}
\label{fig:poln}
\end{center}
\end{figure}

Polarisation images were made for all epochs, except epoch 14 (July
22), when the polarisation angle calibrator 3C\,286 was not observed.
Table~\ref{tab:poln} details the polarisation properties of the
detected components.  Since the overall flux density of the source was
relatively lower during the July observations, polarisation was not
detected at a significant level in the SE component after epoch 12,
which itself only had a $2\sigma$ detection, and no significant core
polarisation was detected at any time.  The polarisation maps are
shown in Fig.\,\ref{fig:poln}, and typically have a r.m.s.\ noise
level of $\sim150\mu$Jy\,beam$^{-1}$.  The SE component appeared to be
significantly polarised, at a level of between 5 and 20 per cent.  In
epochs 1 and 9, the core was also found to be polarised, and in epoch
4 there was marginal evidence for the detection of polarisation in the
emerging receding (NW) component.

\begin{table}
\begin{center}
\caption{Polarisation parameters for the 2001 observations of \grs.
  $P$ is the polarised flux density, $P/I$ is the fractional
  polarisation, and P.A. is the polarisation position angle.  Where
  there was no detection, the $3\sigma$ upper limits have been
  given.}
\begin{tabular}{cccccccc} \hline \hline
Epoch & Component & $P$ (mJy) & $P/I$ & P.A.\\
\hline
1 & Core & $0.67\pm0.08$ & $0.017\pm0.002$ & $54.6\pm5.1$\degr\\
1 & SE & $5.80\pm0.09$ & $0.184\pm0.004$ & $-67.0\pm0.7$\degr\\
2 & SE & $3.59\pm0.10$ & $0.170\pm0.005$ & $-65.8\pm1.1$\degr\\
3 & SE & $1.63\pm0.09$ & $0.086\pm0.005$ & $-51.9\pm2.3$\degr\\
4 & SE & $1.50\pm0.13$ & $0.143\pm0.015$ & $-49.8\pm3.6$\degr\\
4 & NW & $0.65\pm0.12$ & $0.236\pm0.056$ & $-82.0\pm7.2$\degr\\
5 & & $<0.32$ & &\\
6 & & $<0.43$ & &\\
7 & SE3 & $5.14\pm0.09$ & $0.121\pm0.003$ & $-61.9\pm0.8$\degr\\
8 & SE3 & $3.54\pm0.10$ & $0.134\pm0.004$ & $-59.6\pm1.2$\degr\\
9 & Core & $3.52\pm0.16$ & $0.037\pm0.002$ & $73.1\pm5.1$\degr\\
10 & SE & $2.79\pm0.11$ & $0.083\pm0.007$ & $-90.2\pm3.8$\degr\\
11 & SE & $0.63\pm0.09$ & $0.070\pm0.010$ & $-77.8\pm5.8$\degr\\
12 & SE & $0.31\pm0.16$ & $0.119\pm0.066$ & $-74.7\pm15.1$\degr\\
13 & & $<0.49$ & &\\
15 & & $<0.49$ & &\\
16 & & $<0.53$ & &\\
\hline \hline
\end{tabular}
\label{tab:poln}
\end{center}
\end{table}

\begin{figure}
\begin{center}
\includegraphics[width=0.45\textwidth,angle=0,clip=]{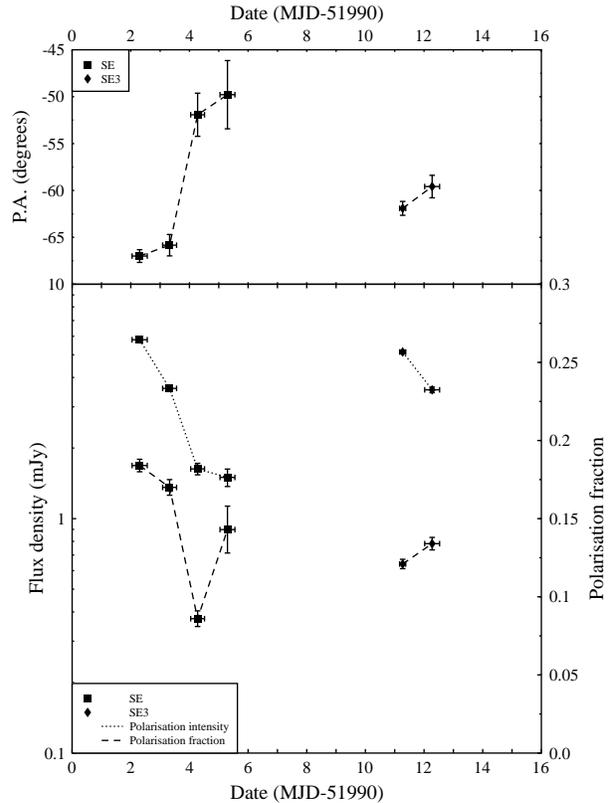}
\caption{Linear polarisation parameters for the 2001 March outburst.
  The bottom panel shows the total intensity on the left-hand axis
  (dotted lines) and the fractional polarisation (dashed lines) on the
  right-hand axis.  The top panel shows the polarisation position
  angle.  The parameters measured for the SE and SE3 components
  are shown by squares and diamonds respectively.}
\label{fig:march_poln}
\end{center}
\end{figure}

The polarisation intensity, fractional polarisation, and polarisation
position angle for the SE and SE3 components are plotted in
Fig.~\ref{fig:march_poln}.  As the component moves outwards, the
polarisation position angle rotates, and both the integrated and
fractional polarisation decrease.  If the change in position angle is
due to changing Faraday rotation as the jet component expands and
moves outwards, the implied changes in rotation measure for the SE and
SE3 components of the March observations are 83.3~rad\,m$^{-2}$ over
the course of 4 days and 11.1~rad\,m$^{-2}$ over 2 days respectively,
whereas the change between epochs 10 and 11 of the July observations is
60.1~rad\,m$^{-2}$ over 2 days.
Without multi-frequency observations, it is not possible to tell
whether the observed rotation of the position angle is in fact due to
Faraday rotation.  But we note that \citet{Fen02} observed a `rotator'
event in 2001 January, when the polarisation position angles at 4.8
and 8.4~GHz rotated together, implying that Faraday rotation was not
responsible.  A different mechanism is therefore required at times in
this source.  Such a mechanism could be that the observed jet knot is
composed of a series of shocks, which fade at different rates, or even
reshock and brighten.  As different shocks come to dominate, the
observed position angle of the polarisation vector would appear to
change if the field lines were oriented differently in the different
shocks.  A more exciting possibility however is that we could be
seeing genuine rotation of the field lines, either tracing an
underlying helical pattern in the field lines at large distances from
the core \citep[e.g.][]{Gom01}, or possibly even due to rotation of
the knots themselves.  But with the limited sampling and single
frequency monitoring available, we cannot distinguish between these
possibilities.

\subsection{Circular polarisation}
\label{sec:circ_pol}
The ATCA observations of \citet{Fen02} found circular
polarisation of $-0.56\pm0.07$~mJy ($0.32\pm0.05$ per cent) at 4.8~GHz just
prior to our observations of epoch 1.  Our MERLIN observations put a
$3\sigma$ upper limit of 0.5~mJy\,beam$^{-1}$ on the Stokes \textit{V}
flux density during epoch 1 when the source was at its brightest,
although the circular polarisation feeds on the MERLIN antennas make
Stokes \textit{V} more susceptible to contamination by Stokes
\textit{I} than is the case for the linearly polarised feeds used on
ATCA.  The ATCA detections were interpreted as either intrinsic
circular polarisation of the synchrotron emission or the conversion of
linear to circular polarisation in the synchrotron-emitting plasma.
In either case, a low-energy tail to the electron energy distribution
is implied.

\section{Energetics}
\label{sec:energetics}
\begin{figure}
\begin{center}
\includegraphics[width=0.45\textwidth,angle=0,clip=]{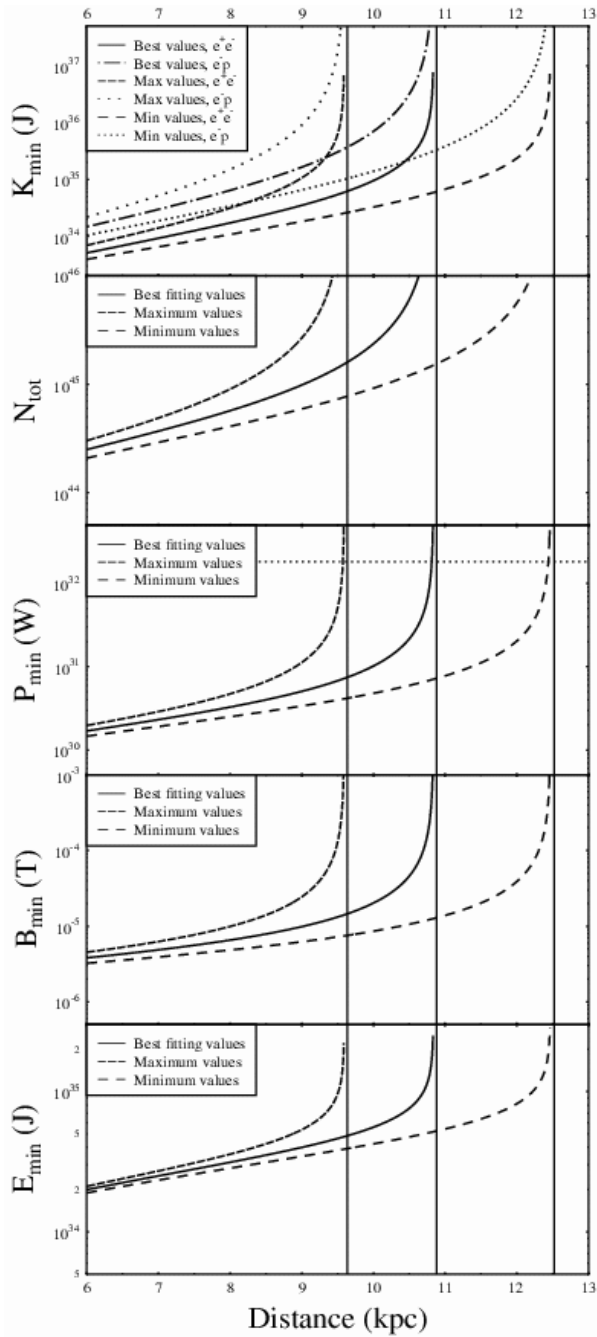}
\caption{Variation with assumed source distance of minimum energy
  ($E_{\rm min}$), minimum energy magnetic field ($B_{\rm min}$),
  minimum power requirement ($P_{\rm min}$), total number of
  relativistic electrons ($N_{\rm tot}$), and minimum kinetic energy
  of jet knot ($K_{\rm min}$).  Calculations have been done for the
  best fitting proper motions, the proper motions both low by
  $1\sigma$ (minimum values), and the proper motions both high by
  $1\sigma$ (maximum values).  The vertical lines indicate the
  corresponding values of $d_{\rm max}$ for these cases, 10.9, 12.5,
  and 9.5~kpc respectively.  The dotted line in the minimum power plot
  shows the Eddington luminosity for a 14$M_{\odot}$ black hole.}
\label{fig:energetics}
\end{center}
\end{figure}

Taking the rise time of the flare observed during epoch 6, $\Delta t$,
as a constraint on the volume of the emitting region, $V=4\pi(c\Delta
t)^3 /3$, and assuming that the increase in flux density originates
from within this region, then using the formalism of \citet{Lon94}, we
can estimate a minimum energy associated with this flare.
\begin{equation}
W_{\rm min} = 3.0\times10^6 \eta^{4/7} 
\left(\frac{V}{\rm{m}^{-3}}\right)^{3/7} 
\left(\frac{\nu}{{\rm Hz}}\right)^{2/7} 
\left(\frac{L_{\nu}}{{\rm W Hz}^{-1}}\right)^{4/7} {\rm J},
\end{equation}
where $V$ is the source volume, $\nu$ is the frequency at which the
luminosity $L_{\nu}$ is measured, and $(\eta-1)$ is the ratio of
energy in protons to that in relativistic electrons.  The magnetic
field corresponding to this minimum energy criterion (close to but not
identical to the equipartition magnetic field), $B_{\rm min}$, may be
expressed \citep{Lon94} as
\begin{equation}
B_{\rm min} = 1.8 \left(\frac{\eta L_{\nu}}{V}\right)^{2/7}
\nu^{1/7}\qquad {\rm T}.
\end{equation}
In order to evaluate these quantities, the rise time of the outburst,
the frequency and the luminosity must first be calculated in the rest
frame of the source.  Assuming that the flux density is dominated by
the approaching jet, the rise time $\Delta t$ in the rest frame of the
observer is Doppler compressed compared to that in the source frame,
$\tau$, as
\begin{equation}
\tau = \delta_{\rm app} \Delta t =
\frac{\Delta t}{\Gamma(1-\beta\cos\theta)},
\end{equation}
The rest frame frequency and flux density may also be found for the
approaching component.  $\nu = \delta_{\rm app}\nu^{\prime}$, where
primed quantities denote those in the rest frame of the jet knot, and
unprimed quantities those in the observer's frame.  Since
$(S_{\nu}/\nu^3)$ is a Lorentz invariant,
\begin{equation}
S_{\nu}^{\prime} = \frac{S_{\nu}}{\delta_{\rm app}^{3+\alpha}} =
S_{\nu}\left(\Gamma[1-\beta\cos\theta]\right)^{3+\alpha},
\end{equation}
where the extra power of $\alpha$ corrects for a source spectrum not
being flat.  The monochromatic luminosity is expressed as $L_{\nu} =
4\pi d^2 S_{\nu}^{\prime}$.  Since this depends on the distance to the
source, the inferred minimum energy and minimum energy magnetic field
are plotted as a function of source distance in
Fig.~\ref{fig:energetics}.  These are derived under the assumptions
that the filling factor of the source is $\sim 1$, that the spectral
index of the emission is $\alpha=0.75$ (consistent with the spectral
index for the integrated emission derived for this outburst by
\citet{Fen02}), that the radio emission extends over a range in
frequency such that $\nu_{\rm max}^{-(p-2)/2} \ll \nu_{\rm
min}^{-(p-2)/2}$, and that the observing frequency $\nu = \nu_{\rm
min}$, the lowest frequency down to which radio emission is seen.
This puts a lower limit on the minimum energy of the outburst.  The
low-energy tail to the electron energy distribution implied by the
circular polarisation observations outlined in \S~\ref{sec:circ_pol}
would however significantly increase the total energy requirement.

The minimum energy is a more slowly-varying function of source
distance than the minimum power, $2 E_{\rm min}/\delta_{\rm app}\Delta
t$ (the factor of two arising from the assumption of symmetric
approaching and receding components), since the Doppler factor
$\delta_{\rm app}$ decreases rapidly as the source distance approaches
$d_{\rm max}$ (see Fig.\,\ref{fig:jetparms}).

Knowing the luminosity of the source (in its rest frame) and the
minimum energy field, we can estimate the total number of relativistic
electrons.  For a power-law electron spectrum, $N(E) = \kappa
E^{-p}\,dE$, assuming that the maximum electron Lorentz factor is
large, we can express
\begin{equation}
\kappa = (p-1)n_{\rm tot}(m_{\rm e}c^2)^{p-1}\gamma_{\rm min}^{p-1},
\end{equation}
where $n_{\rm tot}$ is the electron number density, $m_{\rm e}$ is the
electron mass, and $\gamma_{\rm min}$ the minimum electron Lorentz
factor.  We assume that each electron radiates at frequency
\begin{equation}
\nu = \frac{\gamma^2 e B}{2 \pi m_{\rm e}},
\label{eq:freq}
\end{equation}
where $e$ is the charge of an electron.  Assuming that the minimum
Lorentz factor electrons correspond to the observing frequency, we can
then use equation (19.17) from \citet{Lon94} to derive
\begin{equation}
N_{\rm tot} = \frac{L_{\nu}}{B}\left[A(\alpha)(p-1)(m_{\rm
  e}c^2)^{p-1} \left(\frac{2\pi m_{\rm
  e}}{e}\right)^{\alpha}\right]^{-1},
\end{equation}
where $N_{\rm tot}$ is the total number of relativistic electrons.
$A(\alpha)$ is given by \citep[][p.\ 292]{Lon94}, and is equal to 594
for $\alpha = 0.75$.

As explained by \citet{Fen99}, the kinetic energy associated with the
bulk motion of the jet knot is given by $(\Gamma-1)E_{\rm min}$ for an
electron-positron jet, or by $(\Gamma-1)(E_{\rm min}+N_{\rm tot}m_{\rm
p}c^2)$ for an electron-proton jet with `cold' protons, where
$m_{\rm p}$ is the proton mass.  These two cases are plotted in the top
panel of Fig.~\ref{fig:energetics}, and, unless $d$ is very close to
$d_{\rm max}$, differ only by a factor 2--3.

The uncertainties in the distance set out in \S\,\ref{sec:distance}
should be borne in mind when considering the values in
Fig.~\ref{fig:energetics}.  The minimum energy, minimum power, and
kinetic energy are all lower limits, which would rise if the source
was far from equipartition, which is certainly possible given that the
jet knot is decaying and expanding, and is thus manifestly not in a
steady state.  Furthermore, all these estimates would be further
increased if, as seems to be the case from the detection of circular
polarisation, the electron energy distribution carries on as a
power-law down to energies significantly lower than those associated
with the observing frequency via Equation~\ref{eq:freq}.

\section{Proper motion discrepancy}
\label{sec:previous}
These observations seem to have confirmed the observed discrepancy
between the measured proper motions on milliarcsecond scales, as
measured by the VLBA and MERLIN, and those measured on arcsecond
scales with the VLA.  The proper motions for the SE component of the
1997 October flare found by \citet{Dha00b} over the course of 2.5~h
correspond to $22.1\pm1.9$~mas\,d$^{-1}$, consistent with the value of
$23.6\pm0.5$~mas\,d$^{-1}$ found with MERLIN for the same event
\citep{Fen99}, and imply no deceleration between 50 and 320\,mas.
\citet{Dha00b} also imaged discrete ejecta for an event in 1998 May,
finding a proper motion of $22.3\pm1.7$~mas d$^{-1}$ for the
approaching component over the course of 4.5~h.  However,
\citet{Rod99} consistently measured proper motions of
$\sim17$~mas\,d$^{-1}$ with the VLA for four different outbursts in
1994 January--April \citep[including that analysed by][]{Mir94}.  They
saw no evidence for deceleration between 80 and 1420\,mas.  In this
paper, we again measure higher proper motions with MERLIN on scales of
50--300~mas, with no evidence for deceleration.

The discrepancy between the VLA and MERLIN measurements was attributed
by \citet{Fen99} either to intrinsic differences in the jet speeds for
the two outbursts, or to resolution effects between the two arrays,
whereas \citet{Rod99} also suggested a possible change in the angle of
ejection.  Our observations suggest that the jet velocity does not
vary dramatically between outbursts, since the proper motions measured
for both the 2001 March and July flares were consistent with one
another, and with those measured by \citet{Fen99} and \citet{Dha00b}.
However, we note that with such a small sample size, this conclusion
cannot be regarded as definitive.

As well as the differences in angular resolution between these two
seemingly discrepant sets of results, there is also a difference in
the observation dates.  All the VLA observations were taken in in
1994, while the MERLIN and VLBA data were taken in 1997, 1998 and
2001.  This raises the possibility that the discrepancy could be a
result of jet precession.  If the jet angle had precessed towards us
between the 1994 and the 1997 observations, the measured proper
motions of the approaching component would be greater owing to
relativistic effects.  However, those of the receding component would
then be correspondingly lower \citep[see fig.~4 of][]{Fen03}, in
contradiction to the observations.  The VLA measured receding proper
motions of 7--9\,mas d$^{-1}$, whereas the MERLIN observations give
10--12~mas d$^{-1}$.  Also, we might expect to observe a similar
change between the 1997 and 2001 observations as seen between 1994 and
1997, but the former two results are in fact consistent with one
another.  These arguments would seem to rule out the precession
theory, and lend weight to the suggestion of \citet{Fen99} whereby the
lower angular resolution of the VLA resulted in the blending of
components and hence lower proper motions.  If this is the true
explanation however, it begs the question of whether MERLIN itself is
resolving the components adequately, and whether with further improved
resolution, one would measure still faster proper motions.  Certainly
none of the MERLIN observations show closely-spaced components that
might be blended at the VLA resolution of 200~mas.  That all the VLA
observations should consist of such superpositions of ejecta, whereas
none of the MERLIN observations did seems to be too much of a
coincidence.  Until this issue is resolved, the derived jet speeds
should be considered open to possible future revision upwards, making
this source even more energetic than is currently believed.
Observations of further outbursts with multiple arrays seem to be
required to properly resolve this issue.

A different explanation could be provided by the environment of \grs,
and the medium into which the jet knots propagated.  The source was
originally discovered in 1992 August \citep{Cas92} and has not since
returned to full quiescence.  BATSE was launched in 1991 April, and
\citet{Pac96} state that no flares were detected prior to 1992 May 6.
If prior to its discovery in 1992 the source had been in its true
quiescent state with a very weak jet, the environment of the source
would have been the relatively dense, low-velocity interstellar medium
(ISM).  An outburst at this time would have quickly run into the ISM,
and the interaction between the jet material and the ISM would have
determined the advance speed of the external shocks thus generated.
We note that the infrared jets in \grs\ detected by Sams, Eckart \&
Sunyaev (1996), which have never since been detected, could correspond
to the heating of the surrounding ISM by the jet.  As the jet/ISM
boundary moved further out over time, the jet would have been less
energetic at the boundary, and unable to heat the surrounding ISM to
detectable levels.  Prior to later outbursts, the plateau-state steady
jet could have inflated a bubble in the surrounding ISM similar to
that seen in Cygnus X-1 (Gallo et al., in prep.), evacuating the
region through which the jet propagates.  Consequently, in later
outbursts we see only the emission from the internal shocks as the
high Lorentz factor jet catches up with the lower-velocity steady jet,
since the jet/ISM boundary is considerably more distant from the core
of the system.  Future VLA observations should therefore measure the
same proper motions as those measured by MERLIN and the VLBA.  This
scenario would imply that the lobes identified by \citet{Kai04} are
unlikely to be physically associated with \grs, strengthening the case
for it to lie at a distance of $\sim 11$~kpc.  We reiterate the need
for further VLA observations of the relativistic ejecta during an
outburst of \grs.

\section{Jet ejection}
\label{sec:model}

\citet{Vad03} suggested that the observed flares of \grs\ with
discrete `superluminal' ejecta occurred after a radio-loud hard X-ray
state (the C$_{\rm RL}$ state) in which a continuous flat-spectrum jet
existed in the radio pre-flare plateau state.  They postulated that an
X-ray dip into state A corresponded to the ejection of the
Comptonising cloud in the core of the system at higher velocity than
that of the pre-existing continuous jet, creating an internal shock in
the flow which was observed as discrete, fast-moving, radio-emitting
ejecta.  This model is supported by observational evidence, since
Gallo, Fender \& Pooley (2003) had previously noted that steady jet
outflows for generic black hole X-ray binaries were significantly less
relativistic than transient outbursts, whereas in the specific case of
\grs, lower limits on the speed of the compact jet in the plateau
state were found by \citet{Dha00b} and Rib\'o, Dhawan \& Mirabel
(2004), who measured speeds of $\beta=0.1$ and 0.3--0.4 respectively.

\citet{Fen97} made the connection between infrared and radio flares,
suggesting that the two had a common origin in synchroton emission
from a jet.  \citet{Eik98} studied the simultaneous infrared and X-ray
behaviour of the source, and found a close link between X-ray and
infrared flares, including evidence that an X-ray precursor spike was
associated with the start of a flare.  Finally, \citet{Mir98} made
simultaneous X-ray, infrared and radio observations of \grs\ at two
epochs in 1997, and found that the infrared/radio flares were
associated with hard X-ray dips, also suggesting that the ejection
event occurred at the time of an X-ray spike at the end of the
spectral softening.

\citet{Fen04a} created a synthesis of the above pictures, together
with other work, to suggest that above a certain X-ray hardness, the
source produces a steady outflow.  As the X-ray spectrum softens, the
jet velocity increases monotonically until it crosses some `jet-line',
where the soft X-ray flux peaks.  At this point, the jet velocity
increases rapidly before the jet is shut off, and the faster-moving
material causes internal shocks as it collides with the pre-existing
flow.  These shocks then appear as discrete relativistic ejecta with
high bulk Lorentz factors, such as those we have observed.  \grs\
appears to be continually crossing and recrossing this jet-line,
producing the oscillation events such as those reported by
\citet{Poo97} as the jet switches on and off.  If the source spends
sufficient time in the X-ray hard state C, there is enough
pre-existing material in the continuous jet for internal shocks to
give rise to the discrete, highly relativistic ejecta seen in radio
observations.  This scenario is discussed in more detail by
\citet{Fen04b}, who discussed the energetics of the internal shock
model, finding that 5--40 per cent of the kinetic energy of the fast
jet could be released in the shock, which increases the energy budget
of \S\,\ref{sec:energetics} by up to an order of magnitude.  A
physical explanation of the above scenario was also proposed, whereby
the inner edge of the accretion disc moves closer to the compact
object as the accretion rate rises.  This is supported by the increase
in the frequency of the low-frequency QPOs \citep[figs.\ 5, 8, and 9
of][]{Vad03}, and the values of the inner disc radius derived from
model-fitting to X-ray spectra.  The escape velocity from this point
therefore rises, increasing the jet Lorentz factor until an internal
shock is formed as the jet switches off, the reason for which is not
yet entirely clear.  Observational evidence for the jet switching off
can be seen in Fig.~\ref{fig:core_flux}, where the core flux density
falls to very low levels in epochs 1 and 6 in the immediate aftermath
of the ejections of components SE and SE3 of the March outburst
sequence.

We see from Fig.~\ref{fig:energetics} that the minimum power involved
in launching the jet is of order 1--10 per cent of the Eddington
luminosity for a $14M_{\odot}$ black hole, unless the true distance to
the source is very close to the calculated $d_{\rm max}$, when it is
even greater.  \citet{Gal03} noted that in low/hard state black holes,
jet suppression occurs at a constant fraction of a few per cent of the
Eddington rate as a source moves into a high/soft state.  Thus it
appears that the jet could be the dominant power output channel during
the last moments before it switches off.

\section{Conclusions}
We have observed two flaring sequences from GRS\,1015+105 in 2001 March
and July.  The March sequence showed multiple ejection events, whereas
only a single pair of ejecta was observed in July.  We measured proper
motions of $21.4\pm2.0$, $24.7\pm1.0$ and $23.8\pm2.8$~mas\,d$^{-1}$
for the three bright approaching components, and $11.8\pm2.0$ and
$11.8\pm3.5$~mas\,d$^{-1}$ for the two detected receding components,
consistent with those found with MERLIN by \citet{Fen99}.  The
$3\sigma$ upper limits on the source distance are 13.9~kpc for the
March outburst and 15.1~kpc for the July event.

We propose a possible explanation of the discrepancy between the
proper motions measured with the VLA, whereby prior to 1992, the
source had been in quiescence, such that the environment surrounding
it had been filled in.  Subsequently, the jet has inflated a cavity
so that internal shocks propagate through the steady jet unaffected by
the external medium.

We have observed linear polarisation arising from the approaching jet
component which decreases with increasing angular separation from the
core, with a very gradual rotation of polarisation position angle with
time.  With the current data, we are unable to distinguish between the
possibilities of Faraday rotation, brightening and fading of different
shocks within the beam, rotation of the knots, or a genuine underlying
helical structure in the magnetic field.

We have demonstrated that the energetics of the system, often
calculated for an assumed source distance, vary substantially with
source distance, which could lie anywhere between 6 and 12.2~kpc
according to current estimates.  However, the minimum power required
for the outburst is 1--10 per cent of the Eddington luminosity for the
14$M_{\odot}$ black hole believed to lie at the centre of the system,
although this may be more if either the source lies very close to
$d_{\rm max}$ or does not satisfy the minimum energy criterion.

The data provide support for the internal shock model proposed by
\citet{Kai00} and refined by \citet{Vad03} and \citet{Fen04b}, whereby
the jet velocity increases rapidly at the start of an outburst, before
shutting off after the Lorentz factor of the ejected material peaks.
The increasing velocity causes internal shocks which light up the
underlying outflow and appear as discrete ejecta moving outwards from
the core with constant velocity.  Shocks are only produced once, on
crossing the `jet line' from a hard to a soft X-ray state; after the
initial ejection, no second set of shocks is seen prior to the source
moving back to the X-ray state C with steady nuclear jet emission.

\section*{Acknowledgments}

MERLIN is operated as a National Facility by the University of
Manchester at Jodrell Bank Observatory on behalf of the Particle
Physics and Astronomy Research Council (PPARC). We would like to thank
Vivek Dhawan, Michael Rupen, Marc Klein-Wolt, and Tom Maccarone for
useful discussions.  We also thank Francis Graham-Smith and Bryan
Anderson for useful comments following a careful perusal of the
manuscript.  DGM$^{\rm c}$C would like to acknowledge receipt of a
PPARC studentship during the period of this research.

\label{lastpage}

\end{document}